\pdfoutput=1
\documentclass[%
 aip,
 jmp,%
 amsmath,amssymb,
 reprint,%
twocolumn,
]{revtex4-1}

\usepackage[version=3]{mhchem} % Formula subscripts using \ce{}

\usepackage{amsmath}
\usepackage{amsfonts}
\usepackage{amssymb}
\usepackage{graphicx}
\usepackage{dcolumn}
\usepackage{bm}
\usepackage{algorithm} 
\usepackage{algpseudocode} 
\usepackage{xr}
\usepackage{hyperref}
\usepackage{braket}
\usepackage{xcolor}

\makeatletter
\newcommand*{\addFileDependency}[1]{% argument=file name and extension
  \typeout{(#1)}
  \@addtofilelist{#1}
  \IfFileExists{#1}{}{\typeout{No file #1.}}
}
\makeatother

\usepackage[capitalize]{cleveref}
\crefname{figure}{Fig.}{Figs.}
\Crefname{figure}{Figure}{Figures}
\crefname{table}{Tab.}{Tabs.}
\Crefname{table}{Table}{Tables}
\crefname{equation}{Eq.}{Eqs.}
\Crefname{equation}{Equation}{Equations}
\crefname{section}{Sec.}{Secs.}
\Crefname{section}{Section}{Sections}
\usepackage{xcolor}

\begin{document}

\author{Max Rossmannek}
\email{oss@zurich.ibm.com}
\altaffiliation{These authors contributed equally to this work.}
\affiliation{IBM Quantum, IBM Research – Zürich, 8803 Rüschlikon, Switzerland}
\affiliation{Department of Chemistry, University of Zürich, Winterthurerstrasse 190, 8057 Zürich, Switzerland}

\author{Fabijan Pavo\v{s}evi\'{c}}
\email{fpavosevic@gmail.com}
\altaffiliation{These authors contributed equally to this work.}
\affiliation{Center for Computational Quantum Physics, Flatiron Institute, 162 5th Ave., New York, 10010  NY,  USA}

\author{Angel Rubio}
\email{angel.rubio@mpsd.mpg.de}
\affiliation{Center for Computational Quantum Physics, Flatiron Institute, 162 5th Ave., New York, 10010  NY,  USA}
\affiliation{Max Planck Institute for the Structure and Dynamics of Matter and
Center for Free-Electron Laser Science \& Department of Physics,
Luruper Chaussee 149, 22761 Hamburg, Germany}
\affiliation{Nano-Bio Spectroscopy Group and European Theoretical Spectroscopy Facility (ETSF), Universidad del Pa\'is Vasco (UPV/EHU), Av. Tolosa 72, 20018 San Sebastian, Spain}

\author{Ivano Tavernelli}
\email{ita@zurich.ibm.com}
\affiliation{IBM Quantum, IBM Research – Zürich, 8803 Rüschlikon, Switzerland}

%Title
\title[]
  {Quantum Embedding Method for the Simulation of Strongly Correlated Systems on Quantum Computers}

%Abstract 
\begin{abstract}
Quantum computing has emerged as a promising platform for simulating strongly correlated systems in chemistry, for which the standard quantum chemistry methods are either qualitatively inaccurate or too expensive. However, due to the hardware limitations of the available noisy near-term quantum devices, their application is currently limited only to small chemical systems. One way for extending the range of applicability can be achieved within the quantum embedding approach. Herein, we employ the projection-based embedding method for combining the variational quantum eigensolver (VQE) algorithm, although not limited to, with density functional theory (DFT). The developed VQE-in-DFT method is then implemented efficiently on a real quantum device and employed for simulating the triple bond breaking process in butyronitrile. The results presented herein show that the developed method is a promising approach for simulating systems with a strongly correlated fragment on a quantum computer. The developments as well as the accompanying implementation will benefit many different chemical areas including the computer aided drug design as well as the study of metalloenzymes with a strongly correlated fragment.
\end{abstract}

\maketitle

%\section{Introduction}
Quantum computing offers a promising path for tackling hard problems in chemistry and physics for which the standard quantum chemistry methods are either qualitatively inaccurate or too expensive~\cite{cao2019quantum,bauer2020quantum}. One such problem is an accurate description of the strong electron correlation that plays a critical role in transition metal chemistry, magnetic molecules, free radicals chemistry, photochemistry, catalysis, and bond breaking processes~\cite{lyakh2012multireference,lischka2018multireference,khedkar2021modern,biz2021strongly}. Because strong electron correlation is characterized by multiple degenerate electronic states, many popular quantum chemistry methods that use a single Slater determinant, such as the Hartree-Fock (HF) method and the Kohn-Sham density functional theory (KS-DFT) method~\cite{Sham65_1133}, are bound to fail to correctly describe the electronic structure for such systems~\cite{burke2012}. Traditionally, accurate descriptions of strongly correlated systems require multireference quantum chemistry methods in which the wave function is written as a weighted sum of multiple electronic configurations~\cite{Szalay2011}. However, the number of electronic configurations grows exponentially with the number of correlated electrons. This renders these multireference methods inapplicable to non-trivially sized molecules. As a consequence, a manual reduction of the system to an active space is required which in turn prohibits the widespread and black-box application of these methods.

An alternative way of solving the electronic Schr\"odinger equation is offered by the recent developments of quantum algorithms~\cite{bauer2020quantum}. This has opened a path for the near exact description of strongly correlated systems using quantum-computational resources that scale polynomially with the number of correlated electrons~\cite{aspuru2005simulated,peruzzo2014variational,cao2019quantum,bauer2020quantum}.
However, the currently available noisy near-term quantum computing devices suffer from rather low physical qubit counts (compared to the demands set out for the simulation of molecular systems) and relatively poor gate fidelities, which necessitates careful design of quantum algorithms that minimize the required quantum resources. The most popular quantum algorithm suitable for the noisy near-term quantum devices is the hybrid quantum-classical variational quantum eigensolver (VQE)~\cite{peruzzo2014variational}, which utilizes the quantum device only for the classically intractable parts of the computation. In particular, it employs a quantum device for the state preparation and the ground-state energy measurement, but uses a classical computer to optimize the wave function parameters in order to minimize the molecular energy. Over the last decade, the VQE algorithm has been implemented on various quantum computing architectures, such as on photonic quantum processors~\cite{peruzzo2014variational}, trapped ions~\cite{shen2017quantum,hempel2018quantum}, and superconducting qubits~\cite{o2016scalable,kandala2017hardware}.

The accuracy and the efficiency of the VQE algorithm depend on the parameterized ansatz of the wave function. In its original proposal, the VQE algorithm employed the unitary coupled cluster with singles and doubles (UCCSD) ansatz~\cite{barkoutsos2018ph}. Unfortunately, the UCCSD ansatz generates deep quantum circuits with a large number of parameters to be optimized, limiting its applicability on noisy near-term quantum devices only to the simplest chemical systems~\cite{o2016scalable}. This has prompted development of more efficient ansatze with lower quantum resource requirements~\cite{kandala2017hardware,ryabinkin2018qubit,romero2018strategies,lee2018generalized,grimsley2019adaptive,matsuzawa2020jastrow,tang2021qubit}. A particularly effective approach is the adaptive VQE~\cite{grimsley2019adaptive} algorithm (referred to as ADAPT-VQE), which instead of relying on a fixed wave function ansatz (such as UCCSD) grows the ansatz iteratively. This leads to a dramatic decrease in the ansatz depth and number of parameters~\cite{grimsley2019adaptive,tang2021qubit}.

Despite much progress in the development and improvement of the VQE algorithm and its variants, their applicability to large molecular systems is still not yet possible. 
Therefore, additional reduction in the requirement of quantum resources can be achieved by means of embedding techniques~\cite{jones2020embedding,rossmannek2021quantum,kirsopp2022quantum}. These rely on the local nature of chemical interactions which allow that only a small, chemically active part of a molecular system is treated at a high level of theory, and the rest of the system (the environment) is treated at a lower level of theory. Therefore, in principle, within the quantum embedding approach, a quantum device can be employed for handling a strongly correlated fragment of a molecular system at a high level, whereas the environment subsystem can be described with standard quantum chemical methods, such as HF or DFT, on a classical computer.

Among the large variation of quantum embedding approaches~\cite{sun2016quantum,jones2020embedding}, the projection-based embedding method~\cite{manby2012simple} is favored for its simplicity and great performance in simulating transition metal catalysis, enzyme catalysis, or in the design of lithium-ion battery electrolytes~\cite{lee2019projection}. A key feature of this method is that it is free of the non-additive kinetic potential rendering it suitable for systems in which the active fragment is covalently bonded to the environment fragment. This is achieved by the level shift projection operator that ensures orthogonality of the occupied orbitals between the active fragment and the environment fragment~\cite{manby2012simple}. Furthermore, the lack of any non-additive kinetic term ensures that the sum of energies of the active and environment fragment is equal to the energy of the full system if both fragments are treated at the same level of theory, therefore this method is also referred to as exact within the given level of approximation~\cite{manby2012simple}. Moreover, this method allows for a seamless high level calculation of the active fragment without any modification of the post-SCF code. Within this approach, the computational saving is achieved by performing a high level calculation on the reduced number of occupied orbitals relevant to the active fragment as well as by truncation of the unoccupied (virtual) orbitals~\cite{bennie2015accelerating,claudino2019simple} that are spatially
distant from the active region. This allows that the computational cost of the active fragment is completely independent from the size of the whole system.

To facilitate the simulation of large molecular systems with a strongly correlated fragment on noisy near-term quantum devices, herein we combine the VQE algorithm as well as its ADAPT-VQE variant with the projection-based DFT embedding method coined as VQE-in-DFT. In the remainder of this work, we demonstrate that this approach shows excellent performance on a real quantum device for the triple bond stretching in butyronitrile (CH$_3$CH$_2$CH$_2$CN). The results are expected to be of similar quality for a wide range of other chemical systems. The developments and calculations demonstrated herein highlight the versatility and numerical efficiency of the VQE-in-DFT embedding method. Moreover, this work paves the path for developments of other quantum embedding schemes that employ quantum devices for handling strongly correlated fragments of molecular systems.

%\section{Theory}

For the foundation of this work we rely on the projection-based wave function-in-DFT (WF-in-DFT) embedding method~\cite{manby2012simple}, where the total KS density matrix, $\boldsymbol{\gamma}$, of the molecular system obtained from KS-DFT is partitioned into an active and environment subsystem, $\boldsymbol{\gamma}_{\text{A}}$ and $\boldsymbol{\gamma}_{\text{B}}$, respectively.
The energy of the active subsystem A embedded into the environment subsystem B by means of an XY-in-DFT calculation is given by~\cite{lee2019projection}
%\begin{widetext}
\begin{equation}
\begin{aligned}
    \label{eqn:XY-in-DFT}
    &E_{\text{XY-in-DFT}}[\boldsymbol{\gamma}_{\text{emb}}^{\text{A}};\boldsymbol{\gamma}^{\text{A}},\boldsymbol{\gamma}^{\text{B}}]=\\\newline
    &E_{\text{XY}}[\boldsymbol{\gamma}_{\text{emb}}^{\text{A}}]+E_{\text{DFT}}[\boldsymbol{\gamma}^{\text{A}}+\boldsymbol{\gamma}^{\text{B}}]-E_{\text{DFT}}[\boldsymbol{\gamma}^{\text{A}}]\\\newline
    &+\text{tr}[(\boldsymbol{\gamma}_{\text{emb}}^{\text{A}}-\boldsymbol{\gamma}^{\text{A}}) \, \boldsymbol{v}_{\text{emb}}[\boldsymbol{\gamma}^{\text{A}},\boldsymbol{\gamma}^{\text{B}}]]+\alpha \, \text{tr}[\boldsymbol{\gamma}_{\text{emb}}^{\text{A}}\mathbf{P}^{\text{B}}]
\end{aligned}
\end{equation}
%\end{widetext}
where $E_{\text{XY}}$ is the energy of the embedded subsystem A treated at the XY level of theory (where XY may stand for HF or DFT method) evaluated at the density matrix, $\boldsymbol{\gamma}_{\text{emb}}^{\text{A}}$, describing said embedded subsystem. In this equation, $\textbf{P}^{\text{B}} = \textbf{S} \bm{\gamma}^{\text{B}} \textbf{S}$ is a projector that enforces the orthogonality of the occupied orbitals between the two subsystems where $\textbf{S}$ is the overlap matrix. Furthermore, $\alpha$ is a scaling parameter that is shifting the orbital energies of subsystem B to large values, and $\boldsymbol{v}_{\text{emb}}[\bm{\gamma}^{\text{A}}, \bm{\gamma}^{\text{B}}] = \textbf{g}[\bm{\gamma}^{\text{A}} + \bm{\gamma}^{\text{B}}] - \textbf{g}[\bm{\gamma}^{\text{A}}]$ is the embedding potential where $\textbf{g}$ includes all two-electron interactions (Coulomb, exchange, and exchange-correlation).
Variation of Eq.~\eqref{eqn:XY-in-DFT} with respect to $\boldsymbol{\gamma}_{\text{emb}}^{\text{A}}$ provides the Fock matrix of the embedded subsystem A
\begin{equation}
    \label{eqn:Fock-A}
    \textbf{F}^{\text{A}} = \textbf{h}^{\text{A-in-B}} + \textbf{g}[\bm{\gamma}_{\text{emb}}^{\text{A}}]
\end{equation}
that is solved self-consistently, where $\textbf{h}^{\text{A-in-B}}$ is the effective one-electron core Hamiltonian defined by
\begin{equation}
    \label{eqn:h-A-in-B}
    \textbf{h}^{\text{A-in-B}} = \textbf{h} + \boldsymbol{v}_{\text{emb}}[\bm{\gamma}^{\text{A}}, \bm{\gamma}^{\text{B}}] + \alpha \textbf{P}^{\text{B}}
\end{equation}
and $\textbf{h}$ is the one-electron core Hamiltonian matrix.

In this work, we use the VQE algorithm as the higher level method in the WF-in-DFT scheme, resulting in
%\begin{widetext}
\begin{equation}
\begin{aligned}
    \label{eqn:VQE-in-DFT}
    &E_{\text{VQE-in-DFT}}[\Psi_{\text{VQE}}^{\text{A}};\boldsymbol{\gamma}^{\text{A}},\boldsymbol{\gamma}^{\text{B}}]=\\\newline
    &E_{\text{VQE}}[\Psi_{\text{VQE}}^{\text{A}}]+E_{\text{DFT}}[\boldsymbol{\gamma}^{\text{A}}+\boldsymbol{\gamma}^{\text{B}}]-E_{\text{DFT}}[\boldsymbol{\gamma}^{\text{A}}]\\\newline
    &+\text{tr}[(\boldsymbol{\gamma}_{\text{emb}}^{\text{A}}-\boldsymbol{\gamma}^{\text{A}}) \, \boldsymbol{v}_{\text{emb}}[\boldsymbol{\gamma}^{\text{A}},\boldsymbol{\gamma}^{\text{B}}]]+\alpha \, \text{tr}[\boldsymbol{\gamma}_{\text{emb}}^{\text{A}}\mathbf{P}^{\text{B}}]
\end{aligned}
\end{equation}
%\end{widetext}
where $\Psi_{\text{VQE}}^{\text{A}}$ is the VQE wave function of the subsystem A which minimizes the following energy functional
\begin{equation}
    \label{eqn:VQE-Energy-Funtional}
     E_{\text{VQE}}=\underset{\theta}{\text{min}} \langle\Psi(\theta)|\hat{H}^{\text{A-in-B}}|\Psi(\theta)\rangle
\end{equation}
and $\theta$ are the wave function parameters that are optimized within the VQE algorithm. Here, $\hat{H}^{\text{A-in-B}}$ is the second-quantized Hamiltonian of subsystem A defined as
\begin{equation}
    \label{eqn:Hamiltonian}
    \hat{H}^{\text{A-in-B}} = h^p_q a^q_p + \frac{1}{2}g^{pq}_{rs}a^{rs}_{pq}
\end{equation}
where $a_{p_1p_2...p_n}^{q_1q_2...q_n}=a_{q_1}^{\dagger}a_{q_2}^{\dagger}...a_{q_n}^{\dagger}a_{p_n}...a_{p_2}a_{p_1}$ are the second-quantized excitation operators written in terms of fermionic creation/annihilation ($a^{\dagger}/a$) operators. Moreover, $h^p_q$ corresponds to a matrix element of the effective one-electron core Hamiltonian defined in Eq.~\eqref{eqn:h-A-in-B}, and $g^{pq}_{rs}=\langle rs|pq \rangle$ is the two-electron Coulomb repulsion tensor element in the basis of the molecular orbitals of subsystem A.\cite{lee2019projection} Throughout this work, indices $i,j,k,l,...$, $a,b,c,d,...$, and $p,q,r,s,...$ denote occupied, unoccupied, and general electronic spin orbitals, respectively.

As discussed earlier, in the original proposal the VQE algorithm employed the UCC ansatz which is given by  
\begin{equation}
    \label{eqn:UCC}
    |\Psi_{\text{UCC}}\rangle=e^{\hat{T}-\hat{T}^\dagger}|0\rangle
\end{equation}
Here, $|0\rangle$ represents the reference wave function, which is usually the HF wave function, and $\hat{T}=\hat{T}_1+\hat{T}_2+\hat{T}_3+...=\theta_{\mu}a^{\mu}$ is the excitation cluster operator that accounts for the correlation effects between the quantized electrons. Moreover, $a^\mu=a_\mu^{\dagger}=\{a_{i}^{a},a_{ij}^{ab},a_{ijk}^{abc},...\}$ is a set of single, double, triple, and higher excitation operators where $\mu$ is the excitation manifold. Truncation of the excitation cluster operator to $\hat{T}=\hat{T}_1+\hat{T}_2=\theta^i_a a_i^a+\frac{1}{4}\theta^{ij}_{ab}a_{ij}^{ab}$ defines the UCCSD ansatz. A drawback of the UCCSD method is its large number of wave function parameters and deep circuit prohibiting its application on noisy near-term quantum devices. To circumvent these restrictions, the ADAPT-VQE method iteratively constructs the wave function ansatz from a pre-defined set of operators~\cite{grimsley2019adaptive}.
As part of this work, we have investigated two kinds of operator pools: the UCCSD fermionic excitation operator pool, $\tau^\mu = -\tau_\mu^\dagger = \{a_i^a-a^i_a,a_{ij}^{ab}-a^{ij}_{ab}\}$ (in the following referred to as f-ADAPT), and the corresponding qubit operator pool~\cite{tang2021qubit} (in the following referred to as q-ADAPT). The latter can be obtained in a straight-forward manner from the former, by mapping the fermionic excitation operators into qubit space~\cite{jordan1993algebraic,bravyi2002fermionic,seeley2012bravyi}, which results in a set of grouped Pauli terms, and subsequently treating each Pauli term as an individual operator in the pool.
This results in a much larger set of operators to choose from but enables more fine-grained control over the constructed ansatz~\cite{grimsley2019adaptive}.

\begin{figure*}[ht]
  \centering
  \includegraphics[width=5.5in]{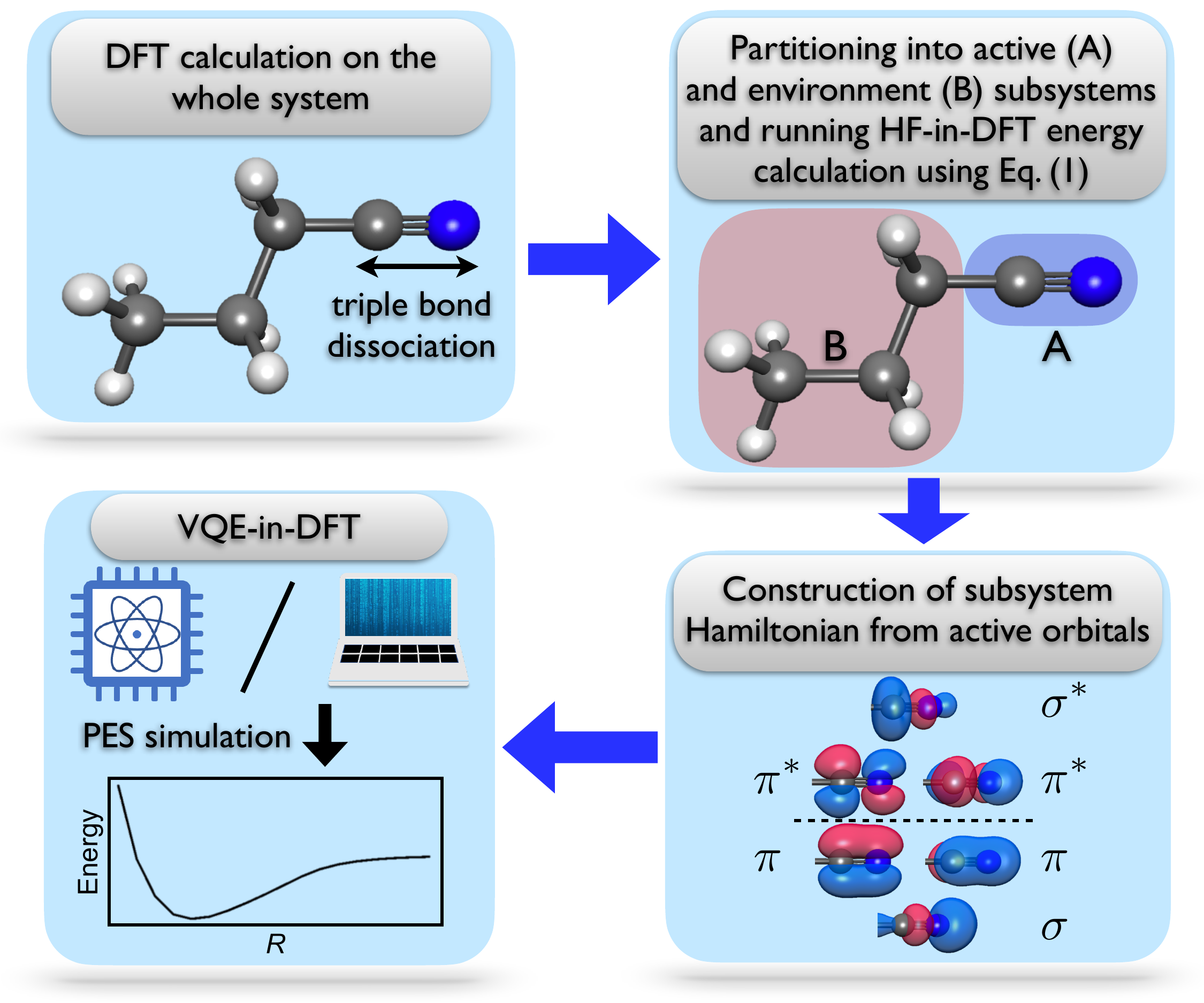}
  \caption{Schematic depiction of the steps performed in this work.}
  \label{fig:scheme}
\end{figure*}

%\section{Results and Discussion}
The VQE-in-DFT method was implemented using \textsf{Qiskit}~\cite{Qiskit_reasonable}. In particular, the VQE-in-HF base implementation will be made publicly available in the \textsf{Qiskit Nature} module~\cite{QiskitNature}. The extension to embedding into DFT will be published separately in combination with the data required to reproduce the results from this work~\cite{SupportingData}. In the following, we study the dissociation of the triple bond in butyronitrile (CH$_3$CH$_2$CH$_2$CN). All calculations were performed at the optimized geometry obtained at CCSD/STO-3G~\cite{hehre1969self} level of theory using the \textsf{Orca} quantum chemistry software~\cite{neese2020orca}. For all further energy calculations we used the STO-3G basis set~\cite{hehre1969self}, while DFT calculations were carried out with the PBE~\cite{perdew1996generalized} exchange-correlation density functional as implemented in an in-house version of the \textsf{Psi4NumPy} quantum chemistry software~\cite{smith2018psi4numpy}.
Under these conditions, the whole system is comprised of 38 electrons in 32 spatial molecular orbitals. The whole molecular system is partitioned into the active strongly correlated fragment \mbox{--CN} and its `environment', \mbox{CH$_3$CH$_2$CH$_2$--}. In the case of VQE-in-HF, the partitioning was carried out with the SPADE procedure~\cite{claudino2019simple}, whereas in the case of VQE-in-PBE, the partitioning was done by means of the combined Pipek-Mezey orbital localization and Mulliken population screening~\cite{manby2012simple}. Truncation of the unoccupied (virtual) orbital space was carried out by means of the one-shell concentric localization of orbitals~\cite{claudino2019simple} procedure employing the same projection basis set as the working basis set (i.e. STO-3G). After the partitioning and truncation of the virtual orbital space, the active fragment is comprised of 14 electrons in 17 spatial molecular orbitals. To further reduce the number of active electrons and orbitals, we have constructed two active spaces; the smaller active space, AS(4,4), contained four electrons in two $\pi$ and two $\pi^*$ CN orbitals, whereas the larger active space, AS(6,6), contained six electrons in two $\pi$, two $\pi^*$, one $\sigma$, and one $\sigma^*$ CN orbitals. Schematic depiction of the workflow for the VQE-in-DFT method is given in Fig.~\ref{fig:scheme}.
In addition to the VQE-in-PBE calculations, we also carried out the FCI-in-PBE and CCSD-in-PBE calculations for comparison.

\begin{figure*}[ht!]
  \centering
  \includegraphics[width=6.5in]{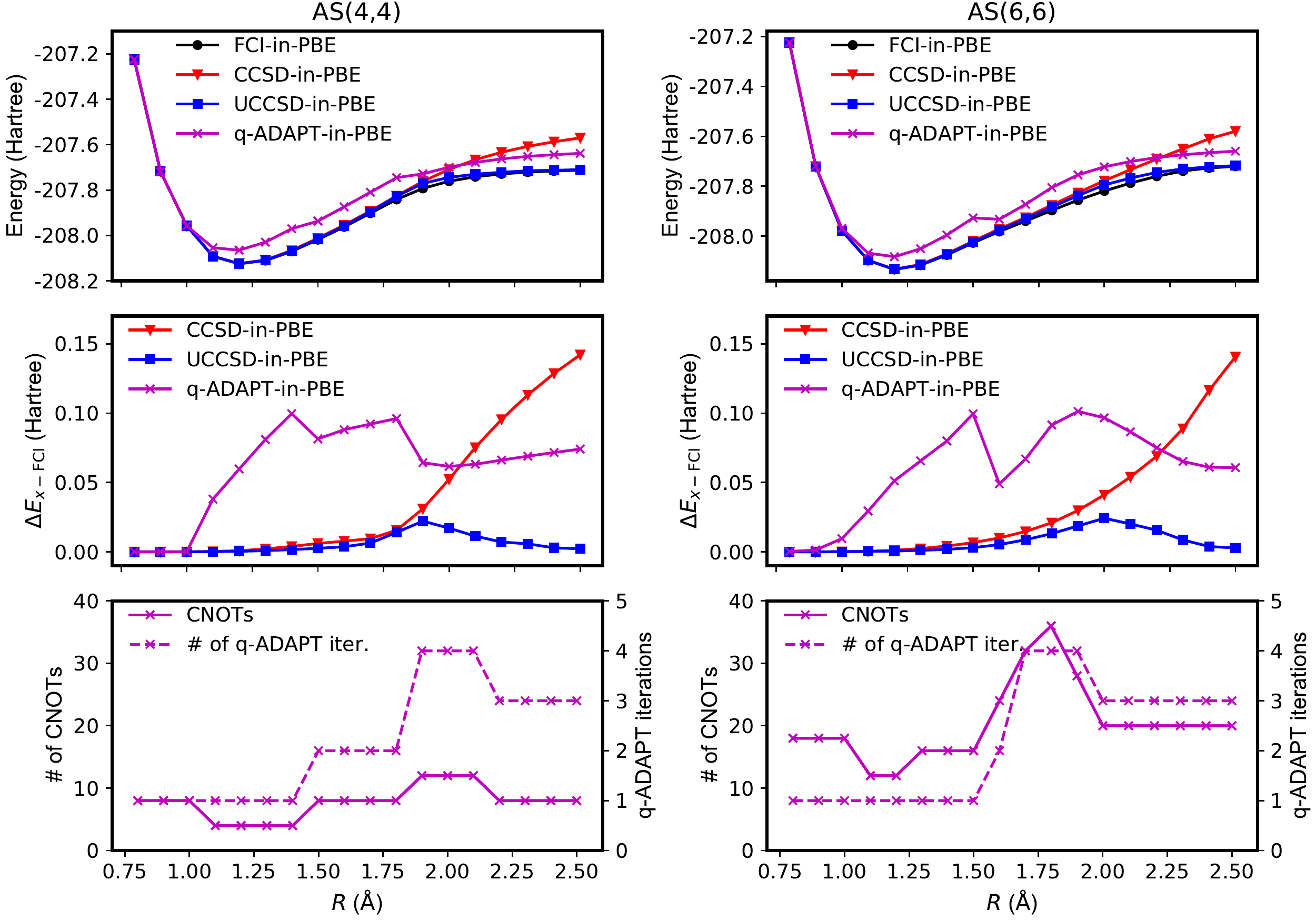}
  \caption{Potential energy surface (upper panels) for the triple C-N bond dissociation in CH$_3$CH$_2$CH$_2$CN calculated with different WF-in-PBE/STO-3G methods and two active spaces, AS(4,4) (left column) and AS(6,6) (right column). The middle panels show the error for a respective embedding method relative to the reference FCI-in-PBE method. The lowest panels indicate the number of CNOT gates (solid magenta line) and number of q-ADAPT-in-PBE iterations (dashed magenta line).}
  \label{fig:in-PBE-PES}
\end{figure*}

In the following, we discuss the results obtained for the VQE-in-PBE simulations. The results obtained with the VQE-in-HF simulations are provided in Section~1 of the Supporting Information (SI).

As preliminary investigation we performed noiseless simulations of the hybrid quantum-classical VQE algorithm. The upper two panels of Fig.~\ref{fig:in-PBE-PES} show the potential energy surface (PES) for the triple C-N bond dissociation in butyronitrile calculated with different WF-in-PBE methods and two active spaces, AS(4,4) (left column) and AS(6,6) (right column). The reference values were calculated with the FCI-in-PBE method (black line) which shows a proper behavior for the triple bond dissociation~\cite{kinoshita2005coupled,cooper2010benchmark}. In case of the CCSD-in-PBE method (red line), the energy curve remains in excellent agreement with the reference curve for values of $R < 1.5~\text{\AA}$, however, for larger values of $R$ the energy starts to deviate rapidly. This trend is even more evident from the middle two panels which show the error of a given method with respect to the reference FCI-in-PBE method. Such deviation in energy for large values of $R$ is not surprising since it is well known that the CCSD method, due to its single-determinant nature, fails in correctly describing the triple bond breaking~\cite{kinoshita2005coupled,cooper2010benchmark}. Equivalent behavior is observed for the PBE-in-PBE (PBE) method and results are provided in Section~2 of the SI.
As opposed to the CCSD-in-PBE and PBE-in-PBE, the UCCSD-in-PBE method (blue line) displays much lower discrepancy in energy with respect to the FCI-in-PBE curve even for large values of $R$ where the strong electron correlation effects are significant. This is again evident from the middle two panels, which show that the blue curve remains nearly flat across all ranges of $R$ values. This behavior is in agreement with the previous findings that the UCCSD method performs well in the case of the triple N-N bond breaking~\cite{grimsley2019trotterized,sokolov2020qUCCSD,rossmannek2021quantum},  which is similar in nature to the triple C-N bond breaking.

Therefore, the UCCSD-in-PBE method is in principle suitable for the simulation of systems with a strongly correlated fragment. However, as already discussed earlier in the text, an implementation of the UCCSD method would require a very deep circuit with a large number of quantum gates, rendering it unsuitable for the application on a noisy near-term quantum computer. 
In particular, the number of CNOT gates required for the implementation of the UCCSD-in-PBE ansatz amounts to 1096 for AS(4,4) and 9200 for AS(6,6), respectively.
It should be noted that these estimates assume an all-to-all connectivity map of physical qubits meaning that the real hardware requirements may change considerably depending on the actual connectivity map and transpiler optimization options.

A more attainable and realistic quantum circuit to be simulated on a noisy near-term quantum device should account to a significantly lower number of CNOT gates. For the application at hand, this was achievable by means of the ADAPT-VQE algorithm. It allows three convergence thresholds to be specified: the maximum number of iterations, the maximum operator gradient, and the maximum change of the energy expectation value between consecutive iterations.
Changing these thresholds allow balancing the ansatz complexity in terms of both, the number of parameters and circuit depth, against the quality of the ansatz.
In Fig.~\ref{fig:in-PBE-PES}, we include the q-ADAPT-in-PBE results obtained for a maximum number of four iterations while targeting a maximum operator gradient of $1\mathrm{e}{-4}$ and a maximum energy expectation value threshold of $100$~mHartree (magenta lines).
In Section~3 of the SI, we provide another set of q-ADAPT-in-PBE results with stricter convergence criteria demonstrating that this method can converge toward the UCCSD-in-PBE results. The q-ADAPT-in-PBE configuration depicted here (magenta lines) was found to strike a good balance of circuit complexity and qualitative dissociation behavior. As shown in the upper two panels of Fig.~\ref{fig:in-PBE-PES}, for $R > 1.0~\text{\AA}$ the energy curve starts to deviate from the reference curve but remains mostly parallel to it, thus, properly describing a correct dissociation of the triple C-N bond for both active spaces. Finally, the lowest panels reveal that for AS(4,4), this choice of convergence thresholds results in at most 12 CNOTs, rendering it suitable for the application on a noisy near-term quantum device.
In contrast, the f-ADAPT-in-PBE method with the same convergence thresholds results in an average number of 115 and 181 CNOTs for the AS(4,4) and AS(6,6), respectively. A selected set of results for the f-ADAPT-in-PBE are given in the Section~4 of the SI.

\begin{figure}[ht!]
  \centering
  \includegraphics[width=3.25in]{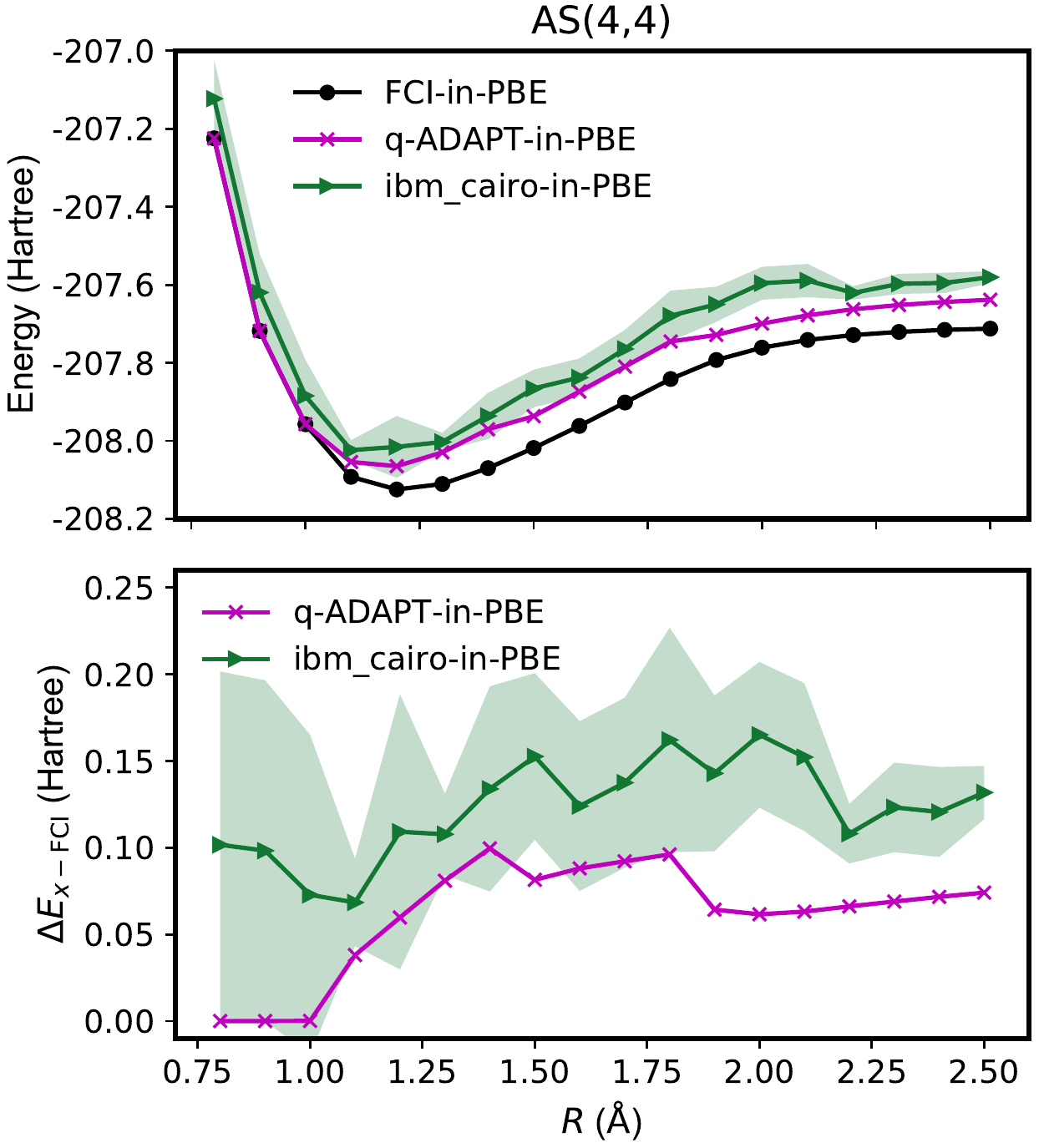}
  \caption{Potential energy surface (upper panel) for the triple C-N bond dissociation in CH$_3$CH$_2$CH$_2$CN calculated with the FCI-in-PBE (black curve), q-ADAPT-in-PBE (magenta line), and  ibm\_cairo-in-PBE (dark green line) methods in the AS(4,4) active space. The shaded green area indicates the standard deviation between 10 independent experiment repetitions and the dark green line corresponds to the average value. The lower panel shows the error for the ibm\_cairo-in-PBE method relative to the reference FCI-in-PBE method.}
  \label{fig:IBM_Cairo_PES}
\end{figure}

To conclude this investigation, we turn our attention to some hardware experiments.  %real device experiments. 
We repeated the final energy evaluation of the q-ADAPT-in-PBE simulations for AS(4,4) on \texttt{ibm\_cairo}, a Falcon r5.11 processor with 27 qubits, using 8192 shots. In the following, we refer to this method as $\text{ibm\_cairo-in-PBE}$.
The Qiskit IBM Runtime service was used via version 0.8.0 of the \texttt{qiskit-ibm-runtime} package.
The quantum circuits were transpiled by the runtime service using optimization level 3 (including dynamical decoupling)~\cite{Qiskit_reasonable}.
Dynamical decoupling was used for error suppression.
Furthermore, the errors were mitigated by means of zero noise extrapolation (ZNE) using a linear extrapolation over noise factors 1, 3, and 5~\cite{youngseok2022zne}.
The noise of all gates was amplified using local gate folding~\cite{LaRose2022mitiqsoftware}.
Finally, the entire ZNE-mitigated measurements were repeated independently ten times. As shown in Fig.~\ref{fig:IBM_Cairo_PES}, the noisy ibm\_cairo-in-PBE results (dark green line) are in a good agreement with respect to the noiseless simulation data (magenta line) across the entire dissociation profile. 
Importantly, the dissociation region corresponding to a large bond distance ($R > 2.2~\text{\AA}$) where the strong correlation is significant is captured well with the proposed quantum embedding algorithm method. The results are also in qualitatively good agreement with the reference FCI-in-PBE data (black line). This is indicative from the lower panel of the same Figure that shows the error with respect to FCI-in-PBE. As can be seen, the noisy results are shifted up by $\sim$120~mHartree relative to the reference FCI-in-PBE curve while remaining parallel across the entire region, thus, properly describing a correct dissociation of the triple C-N bond. This can be seen more clearly in Fig.~S5 where the ibm\_cairo-in-PBE curve has been shifted down by 120~mHartree.

This Letter presents the development and implementation of the projection-embedding VQE-in-DFT method that allows treatment of extended molecular systems on a quantum computer. The developed method is used to study the dissociation of the triple bond in butyronitrile (CH$_3$CH$_2$CH$_2$CN) in which the \mbox{--CN} fragment exhibits a significant multireference character. For assessment of the required quantum resources and accuracy of the developed method, we have performed noiseless simulations by employing the UCCSD ansatz within the VQE algorithm as well as two different variants of the ADAPT-VQE algorithm. We show that the q-ADAPT-in-DFT method, in which the wave function ansatz is adaptively constructed from a pool of qubit operators, strikes a good balance between the number of quantum gates and accuracy. The results presented herein show that the q-ADAPT-in-DFT method is in a qualitatively good agreement with the reference FCI-in-DFT method while keeping the number of CNOT gates below 12 across the entire dissociation profile. The q-ADAPT-in-DFT method is then used for simulation on a noisy near-term quantum device (\texttt{ibm\_cairo}) and the resulting dissociation curve is in good agreement with the noiseless data. 
All together, our results promote the projection embedding scheme as a promising near-term implementation of variational quantum algorithms for electronic structure calculations, bringing this technology one step closer to reaching quantum advantage in quantum chemistry applications, especially for the treatment of systems with a strongly correlated component.
Moreover, the proposed algorithm is of very general applicability;
any future improved quantum algorithm~\cite{aspuru2005simulated} - suitable for error corrected quantum devices - can in fact trivially substitute the VQE optimizer used in this work and bring further benefits. 

In conclusion, we strongly believe that the work presented in this Letter is constituting an important advancement in the development of quantum algorithms for quantum chemistry and will further stimulate the research of improved solutions for near-term and fault-tolerant quantum computers. 
Concerning the specific application, our approach will help shedding new light on the understanding of strongly correlated systems and their essential role in catalysis and the engineering of quantum materials.

\vspace{1em}
\noindent\textbf{Acknowledgements}\\
We acknowledge financial support from the Cluster of Excellence 'CUI: Advanced Imaging of Matter'- EXC 2056 - project ID 390715994 and SFB-925 "Light induced dynamics and control of correlated quantum systems" – project 170620586  of the Deutsche Forschungsgemeinschaft (DFG) and Grupos Consolidados (IT1453-22). We also acknowledge support from the Max Planck–New York Center for Non-Equilibrium Quantum Phenomena. The Flatiron Institute is a division of the Simons Foundation.
This research was supported by the NCCR MARVEL, a National Centre of Competence in Research, funded by the Swiss National Science Foundation (grand number 205602). IBM, the IBM logo, and ibm.com are trademarks of International Business Machines Corp., registered in many jurisdictions worldwide. Other product and service names might be trademarks of IBM or other companies. The current list of IBM trademarks is available at \url{https://www.ibm.com/legal/copytrade}.

\vspace{1em}
\noindent\textbf{Conflict of interest}\\
The authors declare no conflict of interest.

\vspace{1em}
\linespread{1}\selectfont
\noindent\textbf{References}
%\bibliography{references}{}

\begin{thebibliography}{46}%
\makeatletter
\providecommand \@ifxundefined [1]{%
 \@ifx{#1\undefined}
}%
\providecommand \@ifnum [1]{%
 \ifnum #1\expandafter \@firstoftwo
 \else \expandafter \@secondoftwo
 \fi
}%
\providecommand \@ifx [1]{%
 \ifx #1\expandafter \@firstoftwo
 \else \expandafter \@secondoftwo
 \fi
}%
\providecommand \natexlab [1]{#1}%
\providecommand \enquote  [1]{``#1''}%
\providecommand \bibnamefont  [1]{#1}%
\providecommand \bibfnamefont [1]{#1}%
\providecommand \citenamefont [1]{#1}%
\providecommand \href@noop [0]{\@secondoftwo}%
\providecommand \href [0]{\begingroup \@sanitize@url \@href}%
\providecommand \@href[1]{\@@startlink{#1}\@@href}%
\providecommand \@@href[1]{\endgroup#1\@@endlink}%
\providecommand \@sanitize@url [0]{\catcode `\\12\catcode `\$12\catcode
  `\&12\catcode `\#12\catcode `\^12\catcode `\_12\catcode `\%12\relax}%
\providecommand \@@startlink[1]{}%
\providecommand \@@endlink[0]{}%
\providecommand \url  [0]{\begingroup\@sanitize@url \@url }%
\providecommand \@url [1]{\endgroup\@href {#1}{\urlprefix }}%
\providecommand \urlprefix  [0]{URL }%
\providecommand \Eprint [0]{\href }%
\providecommand \doibase [0]{http://dx.doi.org/}%
\providecommand \selectlanguage [0]{\@gobble}%
\providecommand \bibinfo  [0]{\@secondoftwo}%
\providecommand \bibfield  [0]{\@secondoftwo}%
\providecommand \translation [1]{[#1]}%
\providecommand \BibitemOpen [0]{}%
\providecommand \bibitemStop [0]{}%
\providecommand \bibitemNoStop [0]{.\EOS\space}%
\providecommand \EOS [0]{\spacefactor3000\relax}%
\providecommand \BibitemShut  [1]{\csname bibitem#1\endcsname}%
\let\auto@bib@innerbib\@empty
%</preamble>
\bibitem [{\citenamefont {Cao}\ \emph {et~al.}(2019)\citenamefont {Cao},
  \citenamefont {Romero}, \citenamefont {Olson}, \citenamefont {Degroote},
  \citenamefont {Johnson}, \citenamefont {Kieferov{\'a}}, \citenamefont
  {Kivlichan}, \citenamefont {Menke}, \citenamefont {Peropadre}, \citenamefont
  {Sawaya} \emph {et~al.}}]{cao2019quantum}%
  \BibitemOpen
  \bibfield  {author} {\bibinfo {author} {\bibfnamefont {Y.}~\bibnamefont
  {Cao}}, \bibinfo {author} {\bibfnamefont {J.}~\bibnamefont {Romero}},
  \bibinfo {author} {\bibfnamefont {J.~P.}\ \bibnamefont {Olson}}, \bibinfo
  {author} {\bibfnamefont {M.}~\bibnamefont {Degroote}}, \bibinfo {author}
  {\bibfnamefont {P.~D.}\ \bibnamefont {Johnson}}, \bibinfo {author}
  {\bibfnamefont {M.}~\bibnamefont {Kieferov{\'a}}}, \bibinfo {author}
  {\bibfnamefont {I.~D.}\ \bibnamefont {Kivlichan}}, \bibinfo {author}
  {\bibfnamefont {T.}~\bibnamefont {Menke}}, \bibinfo {author} {\bibfnamefont
  {B.}~\bibnamefont {Peropadre}}, \bibinfo {author} {\bibfnamefont {N.~P.}\
  \bibnamefont {Sawaya}},  \emph {et~al.},\ }\bibfield  {title} {\enquote
  {\bibinfo {title} {Quantum chemistry in the age of quantum computing},}\
  }\href@noop {} {\bibfield  {journal} {\bibinfo  {journal} {Chem. Rev.}\
  }\textbf {\bibinfo {volume} {119}},\ \bibinfo {pages} {10856--10915}
  (\bibinfo {year} {2019})}\BibitemShut {NoStop}%
\bibitem [{\citenamefont {Bauer}\ \emph {et~al.}(2020)\citenamefont {Bauer},
  \citenamefont {Bravyi}, \citenamefont {Motta},\ and\ \citenamefont
  {Kin-Lic~Chan}}]{bauer2020quantum}%
  \BibitemOpen
  \bibfield  {author} {\bibinfo {author} {\bibfnamefont {B.}~\bibnamefont
  {Bauer}}, \bibinfo {author} {\bibfnamefont {S.}~\bibnamefont {Bravyi}},
  \bibinfo {author} {\bibfnamefont {M.}~\bibnamefont {Motta}}, \ and\ \bibinfo
  {author} {\bibfnamefont {G.}~\bibnamefont {Kin-Lic~Chan}},\ }\bibfield
  {title} {\enquote {\bibinfo {title} {Quantum algorithms for quantum chemistry
  and quantum materials science},}\ }\href@noop {} {\bibfield  {journal}
  {\bibinfo  {journal} {Chem. Rev.}\ }\textbf {\bibinfo {volume} {120}},\
  \bibinfo {pages} {12685--12717} (\bibinfo {year} {2020})}\BibitemShut
  {NoStop}%
\bibitem [{\citenamefont {Lyakh}\ \emph {et~al.}(2012)\citenamefont {Lyakh},
  \citenamefont {Musia{\l}}, \citenamefont {Lotrich},\ and\ \citenamefont
  {Bartlett}}]{lyakh2012multireference}%
  \BibitemOpen
  \bibfield  {author} {\bibinfo {author} {\bibfnamefont {D.~I.}\ \bibnamefont
  {Lyakh}}, \bibinfo {author} {\bibfnamefont {M.}~\bibnamefont {Musia{\l}}},
  \bibinfo {author} {\bibfnamefont {V.~F.}\ \bibnamefont {Lotrich}}, \ and\
  \bibinfo {author} {\bibfnamefont {R.~J.}\ \bibnamefont {Bartlett}},\
  }\bibfield  {title} {\enquote {\bibinfo {title} {Multireference nature of
  chemistry: The coupled-cluster view},}\ }\href@noop {} {\bibfield  {journal}
  {\bibinfo  {journal} {Chem. Rev.}\ }\textbf {\bibinfo {volume} {112}},\
  \bibinfo {pages} {182--243} (\bibinfo {year} {2012})}\BibitemShut {NoStop}%
\bibitem [{\citenamefont {Lischka}\ \emph {et~al.}(2018)\citenamefont
  {Lischka}, \citenamefont {Nachtigallova}, \citenamefont {Aquino},
  \citenamefont {Szalay}, \citenamefont {Plasser}, \citenamefont {Machado},\
  and\ \citenamefont {Barbatti}}]{lischka2018multireference}%
  \BibitemOpen
  \bibfield  {author} {\bibinfo {author} {\bibfnamefont {H.}~\bibnamefont
  {Lischka}}, \bibinfo {author} {\bibfnamefont {D.}~\bibnamefont
  {Nachtigallova}}, \bibinfo {author} {\bibfnamefont {A.~J.}\ \bibnamefont
  {Aquino}}, \bibinfo {author} {\bibfnamefont {P.~G.}\ \bibnamefont {Szalay}},
  \bibinfo {author} {\bibfnamefont {F.}~\bibnamefont {Plasser}}, \bibinfo
  {author} {\bibfnamefont {F.~B.}\ \bibnamefont {Machado}}, \ and\ \bibinfo
  {author} {\bibfnamefont {M.}~\bibnamefont {Barbatti}},\ }\bibfield  {title}
  {\enquote {\bibinfo {title} {Multireference approaches for excited states of
  molecules},}\ }\href@noop {} {\bibfield  {journal} {\bibinfo  {journal}
  {Chem. Rev.}\ }\textbf {\bibinfo {volume} {118}},\ \bibinfo {pages}
  {7293--7361} (\bibinfo {year} {2018})}\BibitemShut {NoStop}%
\bibitem [{\citenamefont {Khedkar}\ and\ \citenamefont
  {Roemelt}(2021)}]{khedkar2021modern}%
  \BibitemOpen
  \bibfield  {author} {\bibinfo {author} {\bibfnamefont {A.}~\bibnamefont
  {Khedkar}}\ and\ \bibinfo {author} {\bibfnamefont {M.}~\bibnamefont
  {Roemelt}},\ }\bibfield  {title} {\enquote {\bibinfo {title} {Modern
  multireference methods and their application in transition metal
  chemistry},}\ }\href@noop {} {\bibfield  {journal} {\bibinfo  {journal}
  {Phys. Chem. Chem. Phys.}\ } (\bibinfo {year} {2021})}\BibitemShut {NoStop}%
\bibitem [{\citenamefont {Biz}, \citenamefont {Fianchini},\ and\ \citenamefont
  {Gracia}(2021)}]{biz2021strongly}%
  \BibitemOpen
  \bibfield  {author} {\bibinfo {author} {\bibfnamefont {C.}~\bibnamefont
  {Biz}}, \bibinfo {author} {\bibfnamefont {M.}~\bibnamefont {Fianchini}}, \
  and\ \bibinfo {author} {\bibfnamefont {J.}~\bibnamefont {Gracia}},\
  }\bibfield  {title} {\enquote {\bibinfo {title} {Strongly correlated
  electrons in catalysis: Focus on quantum exchange},}\ }\href@noop {}
  {\bibfield  {journal} {\bibinfo  {journal} {ACS Catal.}\ }\textbf {\bibinfo
  {volume} {11}},\ \bibinfo {pages} {14249--14261} (\bibinfo {year}
  {2021})}\BibitemShut {NoStop}%
\bibitem [{\citenamefont {Kohn}\ and\ \citenamefont
  {Sham}(1965)}]{Sham65_1133}%
  \BibitemOpen
  \bibfield  {author} {\bibinfo {author} {\bibfnamefont {W.}~\bibnamefont
  {Kohn}}\ and\ \bibinfo {author} {\bibfnamefont {L.~J.}\ \bibnamefont
  {Sham}},\ }\bibfield  {title} {\enquote {\bibinfo {title} {Self-consistent
  equations including exchange and correlation effects},}\ }\href {\doibase
  10.1103/PhysRev.140.A1133} {\bibfield  {journal} {\bibinfo  {journal} {Phys.
  Rev.}\ }\textbf {\bibinfo {volume} {140}},\ \bibinfo {pages} {A1133--A1138}
  (\bibinfo {year} {1965})}\BibitemShut {NoStop}%
\bibitem [{\citenamefont {Burke}(2012)}]{burke2012}%
  \BibitemOpen
  \bibfield  {author} {\bibinfo {author} {\bibfnamefont {K.}~\bibnamefont
  {Burke}},\ }\bibfield  {title} {\enquote {\bibinfo {title} {Perspective on
  density functional theory},}\ }\href {\doibase 10.1063/1.4704546} {\bibfield
  {journal} {\bibinfo  {journal} {J. Chem. Phys.}\ }\textbf {\bibinfo {volume}
  {136}},\ \bibinfo {pages} {150901} (\bibinfo {year} {2012})},\ \Eprint
  {http://arxiv.org/abs/https://doi.org/10.1063/1.4704546}
  {https://doi.org/10.1063/1.4704546} \BibitemShut {NoStop}%
\bibitem [{\citenamefont {Szalay}\ \emph {et~al.}(2011)\citenamefont {Szalay},
  \citenamefont {M\"{u}ller}, \citenamefont {Gidofalvi}, \citenamefont
  {Lischka},\ and\ \citenamefont {Shepard}}]{Szalay2011}%
  \BibitemOpen
  \bibfield  {author} {\bibinfo {author} {\bibfnamefont {P.~G.}\ \bibnamefont
  {Szalay}}, \bibinfo {author} {\bibfnamefont {T.}~\bibnamefont {M\"{u}ller}},
  \bibinfo {author} {\bibfnamefont {G.}~\bibnamefont {Gidofalvi}}, \bibinfo
  {author} {\bibfnamefont {H.}~\bibnamefont {Lischka}}, \ and\ \bibinfo
  {author} {\bibfnamefont {R.}~\bibnamefont {Shepard}},\ }\bibfield  {title}
  {\enquote {\bibinfo {title} {Multiconfiguration self-consistent field and
  multireference configuration interaction methods and applications},}\ }\href
  {\doibase 10.1021/cr200137a} {\bibfield  {journal} {\bibinfo  {journal}
  {Chem. Rev.}\ }\textbf {\bibinfo {volume} {112}},\ \bibinfo {pages}
  {108--181} (\bibinfo {year} {2011})}\BibitemShut {NoStop}%
\bibitem [{\citenamefont {Aspuru-Guzik}\ \emph {et~al.}(2005)\citenamefont
  {Aspuru-Guzik}, \citenamefont {Dutoi}, \citenamefont {Love},\ and\
  \citenamefont {Head-Gordon}}]{aspuru2005simulated}%
  \BibitemOpen
  \bibfield  {author} {\bibinfo {author} {\bibfnamefont {A.}~\bibnamefont
  {Aspuru-Guzik}}, \bibinfo {author} {\bibfnamefont {A.~D.}\ \bibnamefont
  {Dutoi}}, \bibinfo {author} {\bibfnamefont {P.~J.}\ \bibnamefont {Love}}, \
  and\ \bibinfo {author} {\bibfnamefont {M.}~\bibnamefont {Head-Gordon}},\
  }\bibfield  {title} {\enquote {\bibinfo {title} {Simulated quantum
  computation of molecular energies},}\ }\href@noop {} {\bibfield  {journal}
  {\bibinfo  {journal} {Science}\ }\textbf {\bibinfo {volume} {309}},\ \bibinfo
  {pages} {1704--1707} (\bibinfo {year} {2005})}\BibitemShut {NoStop}%
\bibitem [{\citenamefont {Peruzzo}\ \emph {et~al.}(2014)\citenamefont
  {Peruzzo}, \citenamefont {McClean}, \citenamefont {Shadbolt}, \citenamefont
  {Yung}, \citenamefont {Zhou}, \citenamefont {Love}, \citenamefont
  {Aspuru-Guzik},\ and\ \citenamefont {O’brien}}]{peruzzo2014variational}%
  \BibitemOpen
  \bibfield  {author} {\bibinfo {author} {\bibfnamefont {A.}~\bibnamefont
  {Peruzzo}}, \bibinfo {author} {\bibfnamefont {J.}~\bibnamefont {McClean}},
  \bibinfo {author} {\bibfnamefont {P.}~\bibnamefont {Shadbolt}}, \bibinfo
  {author} {\bibfnamefont {M.-H.}\ \bibnamefont {Yung}}, \bibinfo {author}
  {\bibfnamefont {X.-Q.}\ \bibnamefont {Zhou}}, \bibinfo {author}
  {\bibfnamefont {P.~J.}\ \bibnamefont {Love}}, \bibinfo {author}
  {\bibfnamefont {A.}~\bibnamefont {Aspuru-Guzik}}, \ and\ \bibinfo {author}
  {\bibfnamefont {J.~L.}\ \bibnamefont {O’brien}},\ }\bibfield  {title}
  {\enquote {\bibinfo {title} {A variational eigenvalue solver on a photonic
  quantum processor},}\ }\href@noop {} {\bibfield  {journal} {\bibinfo
  {journal} {Nat. Commun.}\ }\textbf {\bibinfo {volume} {5}},\ \bibinfo {pages}
  {1--7} (\bibinfo {year} {2014})}\BibitemShut {NoStop}%
\bibitem [{\citenamefont {Shen}\ \emph {et~al.}(2017)\citenamefont {Shen},
  \citenamefont {Zhang}, \citenamefont {Zhang}, \citenamefont {Zhang},
  \citenamefont {Yung},\ and\ \citenamefont {Kim}}]{shen2017quantum}%
  \BibitemOpen
  \bibfield  {author} {\bibinfo {author} {\bibfnamefont {Y.}~\bibnamefont
  {Shen}}, \bibinfo {author} {\bibfnamefont {X.}~\bibnamefont {Zhang}},
  \bibinfo {author} {\bibfnamefont {S.}~\bibnamefont {Zhang}}, \bibinfo
  {author} {\bibfnamefont {J.-N.}\ \bibnamefont {Zhang}}, \bibinfo {author}
  {\bibfnamefont {M.-H.}\ \bibnamefont {Yung}}, \ and\ \bibinfo {author}
  {\bibfnamefont {K.}~\bibnamefont {Kim}},\ }\bibfield  {title} {\enquote
  {\bibinfo {title} {Quantum implementation of the unitary coupled cluster for
  simulating molecular electronic structure},}\ }\href@noop {} {\bibfield
  {journal} {\bibinfo  {journal} {Phys. Rev. A}\ }\textbf {\bibinfo {volume}
  {95}},\ \bibinfo {pages} {020501} (\bibinfo {year} {2017})}\BibitemShut
  {NoStop}%
\bibitem [{\citenamefont {Hempel}\ \emph {et~al.}(2018)\citenamefont {Hempel},
  \citenamefont {Maier}, \citenamefont {Romero}, \citenamefont {McClean},
  \citenamefont {Monz}, \citenamefont {Shen}, \citenamefont {Jurcevic},
  \citenamefont {Lanyon}, \citenamefont {Love}, \citenamefont {Babbush} \emph
  {et~al.}}]{hempel2018quantum}%
  \BibitemOpen
  \bibfield  {author} {\bibinfo {author} {\bibfnamefont {C.}~\bibnamefont
  {Hempel}}, \bibinfo {author} {\bibfnamefont {C.}~\bibnamefont {Maier}},
  \bibinfo {author} {\bibfnamefont {J.}~\bibnamefont {Romero}}, \bibinfo
  {author} {\bibfnamefont {J.}~\bibnamefont {McClean}}, \bibinfo {author}
  {\bibfnamefont {T.}~\bibnamefont {Monz}}, \bibinfo {author} {\bibfnamefont
  {H.}~\bibnamefont {Shen}}, \bibinfo {author} {\bibfnamefont {P.}~\bibnamefont
  {Jurcevic}}, \bibinfo {author} {\bibfnamefont {B.~P.}\ \bibnamefont
  {Lanyon}}, \bibinfo {author} {\bibfnamefont {P.}~\bibnamefont {Love}},
  \bibinfo {author} {\bibfnamefont {R.}~\bibnamefont {Babbush}},  \emph
  {et~al.},\ }\bibfield  {title} {\enquote {\bibinfo {title} {Quantum chemistry
  calculations on a trapped-ion quantum simulator},}\ }\href@noop {} {\bibfield
   {journal} {\bibinfo  {journal} {Physical Review X}\ }\textbf {\bibinfo
  {volume} {8}},\ \bibinfo {pages} {031022} (\bibinfo {year}
  {2018})}\BibitemShut {NoStop}%
\bibitem [{\citenamefont {O’Malley}\ \emph {et~al.}(2016)\citenamefont
  {O’Malley}, \citenamefont {Babbush}, \citenamefont {Kivlichan},
  \citenamefont {Romero}, \citenamefont {McClean}, \citenamefont {Barends},
  \citenamefont {Kelly}, \citenamefont {Roushan}, \citenamefont {Tranter},
  \citenamefont {Ding} \emph {et~al.}}]{o2016scalable}%
  \BibitemOpen
  \bibfield  {author} {\bibinfo {author} {\bibfnamefont {P.~J.}\ \bibnamefont
  {O’Malley}}, \bibinfo {author} {\bibfnamefont {R.}~\bibnamefont {Babbush}},
  \bibinfo {author} {\bibfnamefont {I.~D.}\ \bibnamefont {Kivlichan}}, \bibinfo
  {author} {\bibfnamefont {J.}~\bibnamefont {Romero}}, \bibinfo {author}
  {\bibfnamefont {J.~R.}\ \bibnamefont {McClean}}, \bibinfo {author}
  {\bibfnamefont {R.}~\bibnamefont {Barends}}, \bibinfo {author} {\bibfnamefont
  {J.}~\bibnamefont {Kelly}}, \bibinfo {author} {\bibfnamefont
  {P.}~\bibnamefont {Roushan}}, \bibinfo {author} {\bibfnamefont
  {A.}~\bibnamefont {Tranter}}, \bibinfo {author} {\bibfnamefont
  {N.}~\bibnamefont {Ding}},  \emph {et~al.},\ }\bibfield  {title} {\enquote
  {\bibinfo {title} {Scalable quantum simulation of molecular energies},}\
  }\href@noop {} {\bibfield  {journal} {\bibinfo  {journal} {Phys. Rev. X}\
  }\textbf {\bibinfo {volume} {6}},\ \bibinfo {pages} {031007} (\bibinfo {year}
  {2016})}\BibitemShut {NoStop}%
\bibitem [{\citenamefont {Kandala}\ \emph {et~al.}(2017)\citenamefont
  {Kandala}, \citenamefont {Mezzacapo}, \citenamefont {Temme}, \citenamefont
  {Takita}, \citenamefont {Brink}, \citenamefont {Chow},\ and\ \citenamefont
  {Gambetta}}]{kandala2017hardware}%
  \BibitemOpen
  \bibfield  {author} {\bibinfo {author} {\bibfnamefont {A.}~\bibnamefont
  {Kandala}}, \bibinfo {author} {\bibfnamefont {A.}~\bibnamefont {Mezzacapo}},
  \bibinfo {author} {\bibfnamefont {K.}~\bibnamefont {Temme}}, \bibinfo
  {author} {\bibfnamefont {M.}~\bibnamefont {Takita}}, \bibinfo {author}
  {\bibfnamefont {M.}~\bibnamefont {Brink}}, \bibinfo {author} {\bibfnamefont
  {J.~M.}\ \bibnamefont {Chow}}, \ and\ \bibinfo {author} {\bibfnamefont
  {J.~M.}\ \bibnamefont {Gambetta}},\ }\bibfield  {title} {\enquote {\bibinfo
  {title} {Hardware-efficient variational quantum eigensolver for small
  molecules and quantum magnets},}\ }\href@noop {} {\bibfield  {journal}
  {\bibinfo  {journal} {Nature}\ }\textbf {\bibinfo {volume} {549}},\ \bibinfo
  {pages} {242--246} (\bibinfo {year} {2017})}\BibitemShut {NoStop}%
\bibitem [{\citenamefont {Barkoutsos}\ \emph {et~al.}(2018)\citenamefont
  {Barkoutsos}, \citenamefont {Gonthier}, \citenamefont {Sokolov},
  \citenamefont {Moll}, \citenamefont {Salis}, \citenamefont {Fuhrer},
  \citenamefont {Ganzhorn}, \citenamefont {Egger}, \citenamefont {Troyer},
  \citenamefont {Mezzacapo}, \citenamefont {Filipp},\ and\ \citenamefont
  {Tavernelli}}]{barkoutsos2018ph}%
  \BibitemOpen
  \bibfield  {author} {\bibinfo {author} {\bibfnamefont {P.~K.}\ \bibnamefont
  {Barkoutsos}}, \bibinfo {author} {\bibfnamefont {J.~F.}\ \bibnamefont
  {Gonthier}}, \bibinfo {author} {\bibfnamefont {I.}~\bibnamefont {Sokolov}},
  \bibinfo {author} {\bibfnamefont {N.}~\bibnamefont {Moll}}, \bibinfo {author}
  {\bibfnamefont {G.}~\bibnamefont {Salis}}, \bibinfo {author} {\bibfnamefont
  {A.}~\bibnamefont {Fuhrer}}, \bibinfo {author} {\bibfnamefont
  {M.}~\bibnamefont {Ganzhorn}}, \bibinfo {author} {\bibfnamefont {D.~J.}\
  \bibnamefont {Egger}}, \bibinfo {author} {\bibfnamefont {M.}~\bibnamefont
  {Troyer}}, \bibinfo {author} {\bibfnamefont {A.}~\bibnamefont {Mezzacapo}},
  \bibinfo {author} {\bibfnamefont {S.}~\bibnamefont {Filipp}}, \ and\ \bibinfo
  {author} {\bibfnamefont {I.}~\bibnamefont {Tavernelli}},\ }\bibfield  {title}
  {\enquote {\bibinfo {title} {Quantum algorithms for electronic structure
  calculations: Particle-hole hamiltonian and optimized wave-function
  expansions},}\ }\href {\doibase 10.1103/physreva.98.022322} {\bibfield
  {journal} {\bibinfo  {journal} {Phys. Rev. A}\ }\textbf {\bibinfo {volume}
  {98}},\ \bibinfo {pages} {22322} (\bibinfo {year} {2018})}\BibitemShut
  {NoStop}%
\bibitem [{\citenamefont {Ryabinkin}\ \emph {et~al.}(2018)\citenamefont
  {Ryabinkin}, \citenamefont {Yen}, \citenamefont {Genin},\ and\ \citenamefont
  {Izmaylov}}]{ryabinkin2018qubit}%
  \BibitemOpen
  \bibfield  {author} {\bibinfo {author} {\bibfnamefont {I.~G.}\ \bibnamefont
  {Ryabinkin}}, \bibinfo {author} {\bibfnamefont {T.-C.}\ \bibnamefont {Yen}},
  \bibinfo {author} {\bibfnamefont {S.~N.}\ \bibnamefont {Genin}}, \ and\
  \bibinfo {author} {\bibfnamefont {A.~F.}\ \bibnamefont {Izmaylov}},\
  }\bibfield  {title} {\enquote {\bibinfo {title} {Qubit coupled cluster
  method: A systematic approach to quantum chemistry on a quantum computer},}\
  }\href@noop {} {\bibfield  {journal} {\bibinfo  {journal} {J. Chem. Theory
  Comput.}\ }\textbf {\bibinfo {volume} {14}},\ \bibinfo {pages} {6317--6326}
  (\bibinfo {year} {2018})}\BibitemShut {NoStop}%
\bibitem [{\citenamefont {Romero}\ \emph {et~al.}(2018)\citenamefont {Romero},
  \citenamefont {Babbush}, \citenamefont {McClean}, \citenamefont {Hempel},
  \citenamefont {Love},\ and\ \citenamefont
  {Aspuru-Guzik}}]{romero2018strategies}%
  \BibitemOpen
  \bibfield  {author} {\bibinfo {author} {\bibfnamefont {J.}~\bibnamefont
  {Romero}}, \bibinfo {author} {\bibfnamefont {R.}~\bibnamefont {Babbush}},
  \bibinfo {author} {\bibfnamefont {J.~R.}\ \bibnamefont {McClean}}, \bibinfo
  {author} {\bibfnamefont {C.}~\bibnamefont {Hempel}}, \bibinfo {author}
  {\bibfnamefont {P.~J.}\ \bibnamefont {Love}}, \ and\ \bibinfo {author}
  {\bibfnamefont {A.}~\bibnamefont {Aspuru-Guzik}},\ }\bibfield  {title}
  {\enquote {\bibinfo {title} {Strategies for quantum computing molecular
  energies using the unitary coupled cluster ansatz},}\ }\href@noop {}
  {\bibfield  {journal} {\bibinfo  {journal} {Quant. Sci. Techn.}\ }\textbf
  {\bibinfo {volume} {4}},\ \bibinfo {pages} {014008} (\bibinfo {year}
  {2018})}\BibitemShut {NoStop}%
\bibitem [{\citenamefont {Lee}\ \emph {et~al.}(2018)\citenamefont {Lee},
  \citenamefont {Huggins}, \citenamefont {Head-Gordon},\ and\ \citenamefont
  {Whaley}}]{lee2018generalized}%
  \BibitemOpen
  \bibfield  {author} {\bibinfo {author} {\bibfnamefont {J.}~\bibnamefont
  {Lee}}, \bibinfo {author} {\bibfnamefont {W.~J.}\ \bibnamefont {Huggins}},
  \bibinfo {author} {\bibfnamefont {M.}~\bibnamefont {Head-Gordon}}, \ and\
  \bibinfo {author} {\bibfnamefont {K.~B.}\ \bibnamefont {Whaley}},\ }\bibfield
   {title} {\enquote {\bibinfo {title} {Generalized unitary coupled cluster
  wave functions for quantum computation},}\ }\href@noop {} {\bibfield
  {journal} {\bibinfo  {journal} {J. Chem. Theory Comput.}\ }\textbf {\bibinfo
  {volume} {15}},\ \bibinfo {pages} {311--324} (\bibinfo {year}
  {2018})}\BibitemShut {NoStop}%
\bibitem [{\citenamefont {Grimsley}\ \emph
  {et~al.}(2019{\natexlab{a}})\citenamefont {Grimsley}, \citenamefont
  {Economou}, \citenamefont {Barnes},\ and\ \citenamefont
  {Mayhall}}]{grimsley2019adaptive}%
  \BibitemOpen
  \bibfield  {author} {\bibinfo {author} {\bibfnamefont {H.~R.}\ \bibnamefont
  {Grimsley}}, \bibinfo {author} {\bibfnamefont {S.~E.}\ \bibnamefont
  {Economou}}, \bibinfo {author} {\bibfnamefont {E.}~\bibnamefont {Barnes}}, \
  and\ \bibinfo {author} {\bibfnamefont {N.~J.}\ \bibnamefont {Mayhall}},\
  }\bibfield  {title} {\enquote {\bibinfo {title} {An adaptive variational
  algorithm for exact molecular simulations on a quantum computer},}\
  }\href@noop {} {\bibfield  {journal} {\bibinfo  {journal} {Nat. Commun.}\
  }\textbf {\bibinfo {volume} {10}},\ \bibinfo {pages} {1--9} (\bibinfo {year}
  {2019}{\natexlab{a}})}\BibitemShut {NoStop}%
\bibitem [{\citenamefont {Matsuzawa}\ and\ \citenamefont
  {Kurashige}(2020)}]{matsuzawa2020jastrow}%
  \BibitemOpen
  \bibfield  {author} {\bibinfo {author} {\bibfnamefont {Y.}~\bibnamefont
  {Matsuzawa}}\ and\ \bibinfo {author} {\bibfnamefont {Y.}~\bibnamefont
  {Kurashige}},\ }\bibfield  {title} {\enquote {\bibinfo {title} {Jastrow-type
  decomposition in quantum chemistry for low-depth quantum circuits},}\
  }\href@noop {} {\bibfield  {journal} {\bibinfo  {journal} {J. Chem. Theory
  Comput.}\ }\textbf {\bibinfo {volume} {16}},\ \bibinfo {pages} {944--952}
  (\bibinfo {year} {2020})}\BibitemShut {NoStop}%
\bibitem [{\citenamefont {Tang}\ \emph {et~al.}(2021)\citenamefont {Tang},
  \citenamefont {Shkolnikov}, \citenamefont {Barron}, \citenamefont {Grimsley},
  \citenamefont {Mayhall}, \citenamefont {Barnes},\ and\ \citenamefont
  {Economou}}]{tang2021qubit}%
  \BibitemOpen
  \bibfield  {author} {\bibinfo {author} {\bibfnamefont {H.~L.}\ \bibnamefont
  {Tang}}, \bibinfo {author} {\bibfnamefont {V.}~\bibnamefont {Shkolnikov}},
  \bibinfo {author} {\bibfnamefont {G.~S.}\ \bibnamefont {Barron}}, \bibinfo
  {author} {\bibfnamefont {H.~R.}\ \bibnamefont {Grimsley}}, \bibinfo {author}
  {\bibfnamefont {N.~J.}\ \bibnamefont {Mayhall}}, \bibinfo {author}
  {\bibfnamefont {E.}~\bibnamefont {Barnes}}, \ and\ \bibinfo {author}
  {\bibfnamefont {S.~E.}\ \bibnamefont {Economou}},\ }\bibfield  {title}
  {\enquote {\bibinfo {title} {Qubit-adapt-vqe: An adaptive algorithm for
  constructing hardware-efficient ans{\"a}tze on a quantum processor},}\
  }\href@noop {} {\bibfield  {journal} {\bibinfo  {journal} {PRX Quantum}\
  }\textbf {\bibinfo {volume} {2}},\ \bibinfo {pages} {020310} (\bibinfo {year}
  {2021})}\BibitemShut {NoStop}%
\bibitem [{\citenamefont {Jones}\ \emph {et~al.}(2020)\citenamefont {Jones},
  \citenamefont {Mosquera}, \citenamefont {Schatz},\ and\ \citenamefont
  {Ratner}}]{jones2020embedding}%
  \BibitemOpen
  \bibfield  {author} {\bibinfo {author} {\bibfnamefont {L.~O.}\ \bibnamefont
  {Jones}}, \bibinfo {author} {\bibfnamefont {M.~A.}\ \bibnamefont {Mosquera}},
  \bibinfo {author} {\bibfnamefont {G.~C.}\ \bibnamefont {Schatz}}, \ and\
  \bibinfo {author} {\bibfnamefont {M.~A.}\ \bibnamefont {Ratner}},\ }\bibfield
   {title} {\enquote {\bibinfo {title} {Embedding methods for quantum
  chemistry: Applications from materials to life sciences},}\ }\href@noop {}
  {\bibfield  {journal} {\bibinfo  {journal} {J. Am. Chem. Soc.}\ }\textbf
  {\bibinfo {volume} {142}},\ \bibinfo {pages} {3281--3295} (\bibinfo {year}
  {2020})}\BibitemShut {NoStop}%
\bibitem [{\citenamefont {Rossmannek}\ \emph {et~al.}(2021)\citenamefont
  {Rossmannek}, \citenamefont {Barkoutsos}, \citenamefont {Ollitrault},\ and\
  \citenamefont {Tavernelli}}]{rossmannek2021quantum}%
  \BibitemOpen
  \bibfield  {author} {\bibinfo {author} {\bibfnamefont {M.}~\bibnamefont
  {Rossmannek}}, \bibinfo {author} {\bibfnamefont {P.~K.}\ \bibnamefont
  {Barkoutsos}}, \bibinfo {author} {\bibfnamefont {P.~J.}\ \bibnamefont
  {Ollitrault}}, \ and\ \bibinfo {author} {\bibfnamefont {I.}~\bibnamefont
  {Tavernelli}},\ }\bibfield  {title} {\enquote {\bibinfo {title} {Quantum
  hf/dft-embedding algorithms for electronic structure calculations: Scaling up
  to complex molecular systems},}\ }\href@noop {} {\bibfield  {journal}
  {\bibinfo  {journal} {J. Chem. Phys.}\ }\textbf {\bibinfo {volume} {154}},\
  \bibinfo {pages} {114105} (\bibinfo {year} {2021})}\BibitemShut {NoStop}%
\bibitem [{\citenamefont {Kirsopp}\ \emph {et~al.}(2022)\citenamefont
  {Kirsopp}, \citenamefont {Di~Paola}, \citenamefont {Manrique}, \citenamefont
  {Krompiec}, \citenamefont {Greene-Diniz}, \citenamefont {Guba}, \citenamefont
  {Meyder}, \citenamefont {Wolf}, \citenamefont {Strahm},\ and\ \citenamefont
  {Mu{\~n}oz~Ramo}}]{kirsopp2022quantum}%
  \BibitemOpen
  \bibfield  {author} {\bibinfo {author} {\bibfnamefont {J.~J.}\ \bibnamefont
  {Kirsopp}}, \bibinfo {author} {\bibfnamefont {C.}~\bibnamefont {Di~Paola}},
  \bibinfo {author} {\bibfnamefont {D.~Z.}\ \bibnamefont {Manrique}}, \bibinfo
  {author} {\bibfnamefont {M.}~\bibnamefont {Krompiec}}, \bibinfo {author}
  {\bibfnamefont {G.}~\bibnamefont {Greene-Diniz}}, \bibinfo {author}
  {\bibfnamefont {W.}~\bibnamefont {Guba}}, \bibinfo {author} {\bibfnamefont
  {A.}~\bibnamefont {Meyder}}, \bibinfo {author} {\bibfnamefont
  {D.}~\bibnamefont {Wolf}}, \bibinfo {author} {\bibfnamefont {M.}~\bibnamefont
  {Strahm}}, \ and\ \bibinfo {author} {\bibfnamefont {D.}~\bibnamefont
  {Mu{\~n}oz~Ramo}},\ }\bibfield  {title} {\enquote {\bibinfo {title} {Quantum
  computational quantification of protein--ligand interactions},}\ }\href@noop
  {} {\bibfield  {journal} {\bibinfo  {journal} {Int. J. Quant. Chem.}\
  }\textbf {\bibinfo {volume} {122}},\ \bibinfo {pages} {e26975} (\bibinfo
  {year} {2022})}\BibitemShut {NoStop}%
\bibitem [{\citenamefont {Sun}\ and\ \citenamefont
  {Chan}(2016)}]{sun2016quantum}%
  \BibitemOpen
  \bibfield  {author} {\bibinfo {author} {\bibfnamefont {Q.}~\bibnamefont
  {Sun}}\ and\ \bibinfo {author} {\bibfnamefont {G.~K.-L.}\ \bibnamefont
  {Chan}},\ }\bibfield  {title} {\enquote {\bibinfo {title} {Quantum embedding
  theories},}\ }\href@noop {} {\bibfield  {journal} {\bibinfo  {journal} {Acc.
  Chem. Res.}\ }\textbf {\bibinfo {volume} {49}},\ \bibinfo {pages}
  {2705--2712} (\bibinfo {year} {2016})}\BibitemShut {NoStop}%
\bibitem [{\citenamefont {Manby}\ \emph {et~al.}(2012)\citenamefont {Manby},
  \citenamefont {Stella}, \citenamefont {Goodpaster},\ and\ \citenamefont
  {Miller~III}}]{manby2012simple}%
  \BibitemOpen
  \bibfield  {author} {\bibinfo {author} {\bibfnamefont {F.~R.}\ \bibnamefont
  {Manby}}, \bibinfo {author} {\bibfnamefont {M.}~\bibnamefont {Stella}},
  \bibinfo {author} {\bibfnamefont {J.~D.}\ \bibnamefont {Goodpaster}}, \ and\
  \bibinfo {author} {\bibfnamefont {T.~F.}\ \bibnamefont {Miller~III}},\
  }\bibfield  {title} {\enquote {\bibinfo {title} {A simple, exact
  density-functional-theory embedding scheme},}\ }\href@noop {} {\bibfield
  {journal} {\bibinfo  {journal} {J. Chem. Theory Comput.}\ }\textbf {\bibinfo
  {volume} {8}},\ \bibinfo {pages} {2564--2568} (\bibinfo {year}
  {2012})}\BibitemShut {NoStop}%
\bibitem [{\citenamefont {Lee}\ \emph {et~al.}(2019)\citenamefont {Lee},
  \citenamefont {Welborn}, \citenamefont {Manby},\ and\ \citenamefont
  {Miller~III}}]{lee2019projection}%
  \BibitemOpen
  \bibfield  {author} {\bibinfo {author} {\bibfnamefont {S.~J.}\ \bibnamefont
  {Lee}}, \bibinfo {author} {\bibfnamefont {M.}~\bibnamefont {Welborn}},
  \bibinfo {author} {\bibfnamefont {F.~R.}\ \bibnamefont {Manby}}, \ and\
  \bibinfo {author} {\bibfnamefont {T.~F.}\ \bibnamefont {Miller~III}},\
  }\bibfield  {title} {\enquote {\bibinfo {title} {Projection-based
  wavefunction-in-dft embedding},}\ }\href@noop {} {\bibfield  {journal}
  {\bibinfo  {journal} {Acc. Chem. Res.}\ }\textbf {\bibinfo {volume} {52}},\
  \bibinfo {pages} {1359--1368} (\bibinfo {year} {2019})}\BibitemShut {NoStop}%
\bibitem [{\citenamefont {Bennie}\ \emph {et~al.}(2015)\citenamefont {Bennie},
  \citenamefont {Stella}, \citenamefont {Miller~III},\ and\ \citenamefont
  {Manby}}]{bennie2015accelerating}%
  \BibitemOpen
  \bibfield  {author} {\bibinfo {author} {\bibfnamefont {S.~J.}\ \bibnamefont
  {Bennie}}, \bibinfo {author} {\bibfnamefont {M.}~\bibnamefont {Stella}},
  \bibinfo {author} {\bibfnamefont {T.~F.}\ \bibnamefont {Miller~III}}, \ and\
  \bibinfo {author} {\bibfnamefont {F.~R.}\ \bibnamefont {Manby}},\ }\bibfield
  {title} {\enquote {\bibinfo {title} {Accelerating wavefunction in
  density-functional-theory embedding by truncating the active basis set},}\
  }\href@noop {} {\bibfield  {journal} {\bibinfo  {journal} {J. Chem. Phys.}\
  }\textbf {\bibinfo {volume} {143}},\ \bibinfo {pages} {024105} (\bibinfo
  {year} {2015})}\BibitemShut {NoStop}%
\bibitem [{\citenamefont {Claudino}\ and\ \citenamefont
  {Mayhall}(2019)}]{claudino2019simple}%
  \BibitemOpen
  \bibfield  {author} {\bibinfo {author} {\bibfnamefont {D.}~\bibnamefont
  {Claudino}}\ and\ \bibinfo {author} {\bibfnamefont {N.~J.}\ \bibnamefont
  {Mayhall}},\ }\bibfield  {title} {\enquote {\bibinfo {title} {Simple and
  efficient truncation of virtual spaces in embedded wave functions via
  concentric localization},}\ }\href@noop {} {\bibfield  {journal} {\bibinfo
  {journal} {J. Chem. Theory Comput.}\ }\textbf {\bibinfo {volume} {15}},\
  \bibinfo {pages} {6085--6096} (\bibinfo {year} {2019})}\BibitemShut {NoStop}%
\bibitem [{\citenamefont {Jordan}, \citenamefont {Neumann},\ and\ \citenamefont
  {Wigner}(1993)}]{jordan1993algebraic}%
  \BibitemOpen
  \bibfield  {author} {\bibinfo {author} {\bibfnamefont {P.}~\bibnamefont
  {Jordan}}, \bibinfo {author} {\bibfnamefont {J.~v.}\ \bibnamefont {Neumann}},
  \ and\ \bibinfo {author} {\bibfnamefont {E.~P.}\ \bibnamefont {Wigner}},\
  }\bibfield  {title} {\enquote {\bibinfo {title} {On an algebraic
  generalization of the quantum mechanical formalism},}\ }in\ \href@noop {}
  {\emph {\bibinfo {booktitle} {The Collected Works of Eugene Paul Wigner}}}\
  (\bibinfo  {publisher} {Springer},\ \bibinfo {year} {1993})\ pp.\ \bibinfo
  {pages} {298--333}\BibitemShut {NoStop}%
\bibitem [{\citenamefont {Bravyi}\ and\ \citenamefont
  {Kitaev}(2002)}]{bravyi2002fermionic}%
  \BibitemOpen
  \bibfield  {author} {\bibinfo {author} {\bibfnamefont {S.~B.}\ \bibnamefont
  {Bravyi}}\ and\ \bibinfo {author} {\bibfnamefont {A.~Y.}\ \bibnamefont
  {Kitaev}},\ }\bibfield  {title} {\enquote {\bibinfo {title} {Fermionic
  quantum computation},}\ }\href@noop {} {\bibfield  {journal} {\bibinfo
  {journal} {Ann. Phys.}\ }\textbf {\bibinfo {volume} {298}},\ \bibinfo {pages}
  {210--226} (\bibinfo {year} {2002})}\BibitemShut {NoStop}%
\bibitem [{\citenamefont {Seeley}, \citenamefont {Richard},\ and\ \citenamefont
  {Love}(2012)}]{seeley2012bravyi}%
  \BibitemOpen
  \bibfield  {author} {\bibinfo {author} {\bibfnamefont {J.~T.}\ \bibnamefont
  {Seeley}}, \bibinfo {author} {\bibfnamefont {M.~J.}\ \bibnamefont {Richard}},
  \ and\ \bibinfo {author} {\bibfnamefont {P.~J.}\ \bibnamefont {Love}},\
  }\bibfield  {title} {\enquote {\bibinfo {title} {The bravyi-kitaev
  transformation for quantum computation of electronic structure},}\
  }\href@noop {} {\bibfield  {journal} {\bibinfo  {journal} {J. Chem. Phys.}\
  }\textbf {\bibinfo {volume} {137}},\ \bibinfo {pages} {224109} (\bibinfo
  {year} {2012})}\BibitemShut {NoStop}%
\bibitem [{\citenamefont {tA~v}\ \emph {et~al.}(2021)\citenamefont {tA~v},
  \citenamefont {ANIS}, \citenamefont {Abby-Mitchell},\ and\ \citenamefont
  {et~al.}}]{Qiskit_reasonable}%
  \BibitemOpen
  \bibfield  {author} {\bibinfo {author} {\bibfnamefont {A.}~\bibnamefont
  {tA~v}}, \bibinfo {author} {\bibfnamefont {M.~S.}\ \bibnamefont {ANIS}},
  \bibinfo {author} {\bibnamefont {Abby-Mitchell}}, \ and\ \bibinfo {author}
  {\bibfnamefont {H.~A.}\ \bibnamefont {et~al.}},\ }\bibfield  {title}
  {\enquote {\bibinfo {title} {Qiskit: An open-source framework for quantum
  computing},}\ }\href {\doibase 10.5281/zenodo.2573505} {\  (\bibinfo {year}
  {2021}),\ 10.5281/zenodo.2573505}\BibitemShut {NoStop}%
\bibitem [{Qis(2023)}]{QiskitNature}%
  \BibitemOpen
  \href {https://github.com/Qiskit/qiskit-nature} {\enquote {\bibinfo {title}
  {Qiskit {N}ature},}\ } (\bibinfo {year} {2023}),\ \bibinfo {note}
  {https://github.com/Qiskit/qiskit-nature}\BibitemShut {NoStop}%
\bibitem [{Sup(2023)}]{SupportingData}%
  \BibitemOpen
  \href {https://github.com/mrossinek/projection-embedding-vqe} {} (\bibinfo
  {year} {2023})\BibitemShut {NoStop}%
\bibitem [{\citenamefont {Hehre}, \citenamefont {Stewart},\ and\ \citenamefont
  {Pople}(1969)}]{hehre1969self}%
  \BibitemOpen
  \bibfield  {author} {\bibinfo {author} {\bibfnamefont {W.~J.}\ \bibnamefont
  {Hehre}}, \bibinfo {author} {\bibfnamefont {R.~F.}\ \bibnamefont {Stewart}},
  \ and\ \bibinfo {author} {\bibfnamefont {J.~A.}\ \bibnamefont {Pople}},\
  }\bibfield  {title} {\enquote {\bibinfo {title} {self-consistent
  molecular-orbital methods. i. use of gaussian expansions of slater-type
  atomic orbitals},}\ }\href@noop {} {\bibfield  {journal} {\bibinfo  {journal}
  {J. Chem. Phys.}\ }\textbf {\bibinfo {volume} {51}},\ \bibinfo {pages}
  {2657--2664} (\bibinfo {year} {1969})}\BibitemShut {NoStop}%
\bibitem [{\citenamefont {Neese}\ \emph {et~al.}(2020)\citenamefont {Neese},
  \citenamefont {Wennmohs}, \citenamefont {Becker},\ and\ \citenamefont
  {Riplinger}}]{neese2020orca}%
  \BibitemOpen
  \bibfield  {author} {\bibinfo {author} {\bibfnamefont {F.}~\bibnamefont
  {Neese}}, \bibinfo {author} {\bibfnamefont {F.}~\bibnamefont {Wennmohs}},
  \bibinfo {author} {\bibfnamefont {U.}~\bibnamefont {Becker}}, \ and\ \bibinfo
  {author} {\bibfnamefont {C.}~\bibnamefont {Riplinger}},\ }\bibfield  {title}
  {\enquote {\bibinfo {title} {The orca quantum chemistry program package},}\
  }\href@noop {} {\bibfield  {journal} {\bibinfo  {journal} {J. Chem. Phys.}\
  }\textbf {\bibinfo {volume} {152}},\ \bibinfo {pages} {224108} (\bibinfo
  {year} {2020})}\BibitemShut {NoStop}%
\bibitem [{\citenamefont {Perdew}, \citenamefont {Burke},\ and\ \citenamefont
  {Ernzerhof}(1996)}]{perdew1996generalized}%
  \BibitemOpen
  \bibfield  {author} {\bibinfo {author} {\bibfnamefont {J.~P.}\ \bibnamefont
  {Perdew}}, \bibinfo {author} {\bibfnamefont {K.}~\bibnamefont {Burke}}, \
  and\ \bibinfo {author} {\bibfnamefont {M.}~\bibnamefont {Ernzerhof}},\
  }\bibfield  {title} {\enquote {\bibinfo {title} {Generalized gradient
  approximation made simple},}\ }\href@noop {} {\bibfield  {journal} {\bibinfo
  {journal} {Phys. Rev. Lett.}\ }\textbf {\bibinfo {volume} {77}},\ \bibinfo
  {pages} {3865} (\bibinfo {year} {1996})}\BibitemShut {NoStop}%
\bibitem [{\citenamefont {Smith}\ \emph {et~al.}(2018)\citenamefont {Smith},
  \citenamefont {Burns}, \citenamefont {Sirianni}, \citenamefont {Nascimento},
  \citenamefont {Kumar}, \citenamefont {James}, \citenamefont {Schriber},
  \citenamefont {Zhang}, \citenamefont {Zhang}, \citenamefont {Abbott} \emph
  {et~al.}}]{smith2018psi4numpy}%
  \BibitemOpen
  \bibfield  {author} {\bibinfo {author} {\bibfnamefont {D.~G.}\ \bibnamefont
  {Smith}}, \bibinfo {author} {\bibfnamefont {L.~A.}\ \bibnamefont {Burns}},
  \bibinfo {author} {\bibfnamefont {D.~A.}\ \bibnamefont {Sirianni}}, \bibinfo
  {author} {\bibfnamefont {D.~R.}\ \bibnamefont {Nascimento}}, \bibinfo
  {author} {\bibfnamefont {A.}~\bibnamefont {Kumar}}, \bibinfo {author}
  {\bibfnamefont {A.~M.}\ \bibnamefont {James}}, \bibinfo {author}
  {\bibfnamefont {J.~B.}\ \bibnamefont {Schriber}}, \bibinfo {author}
  {\bibfnamefont {T.}~\bibnamefont {Zhang}}, \bibinfo {author} {\bibfnamefont
  {B.}~\bibnamefont {Zhang}}, \bibinfo {author} {\bibfnamefont {A.~S.}\
  \bibnamefont {Abbott}},  \emph {et~al.},\ }\bibfield  {title} {\enquote
  {\bibinfo {title} {Psi4numpy: An interactive quantum chemistry programming
  environment for reference implementations and rapid development},}\
  }\href@noop {} {\bibfield  {journal} {\bibinfo  {journal} {J. Chem. Theory
  Comput.}\ }\textbf {\bibinfo {volume} {14}},\ \bibinfo {pages} {3504--3511}
  (\bibinfo {year} {2018})}\BibitemShut {NoStop}%
\bibitem [{\citenamefont {Kinoshita}, \citenamefont {Hino},\ and\ \citenamefont
  {Bartlett}(2005)}]{kinoshita2005coupled}%
  \BibitemOpen
  \bibfield  {author} {\bibinfo {author} {\bibfnamefont {T.}~\bibnamefont
  {Kinoshita}}, \bibinfo {author} {\bibfnamefont {O.}~\bibnamefont {Hino}}, \
  and\ \bibinfo {author} {\bibfnamefont {R.~J.}\ \bibnamefont {Bartlett}},\
  }\bibfield  {title} {\enquote {\bibinfo {title} {Coupled-cluster method
  tailored by configuration interaction},}\ }\href@noop {} {\bibfield
  {journal} {\bibinfo  {journal} {J. Chem. Phys.}\ }\textbf {\bibinfo {volume}
  {123}},\ \bibinfo {pages} {074106} (\bibinfo {year} {2005})}\BibitemShut
  {NoStop}%
\bibitem [{\citenamefont {Cooper}\ and\ \citenamefont
  {Knowles}(2010)}]{cooper2010benchmark}%
  \BibitemOpen
  \bibfield  {author} {\bibinfo {author} {\bibfnamefont {B.}~\bibnamefont
  {Cooper}}\ and\ \bibinfo {author} {\bibfnamefont {P.~J.}\ \bibnamefont
  {Knowles}},\ }\bibfield  {title} {\enquote {\bibinfo {title} {Benchmark
  studies of variational, unitary and extended coupled cluster methods},}\
  }\href@noop {} {\bibfield  {journal} {\bibinfo  {journal} {J. Chem. Phys.}\
  }\textbf {\bibinfo {volume} {133}},\ \bibinfo {pages} {234102} (\bibinfo
  {year} {2010})}\BibitemShut {NoStop}%
\bibitem [{\citenamefont {Grimsley}\ \emph
  {et~al.}(2019{\natexlab{b}})\citenamefont {Grimsley}, \citenamefont
  {Claudino}, \citenamefont {Economou}, \citenamefont {Barnes},\ and\
  \citenamefont {Mayhall}}]{grimsley2019trotterized}%
  \BibitemOpen
  \bibfield  {author} {\bibinfo {author} {\bibfnamefont {H.~R.}\ \bibnamefont
  {Grimsley}}, \bibinfo {author} {\bibfnamefont {D.}~\bibnamefont {Claudino}},
  \bibinfo {author} {\bibfnamefont {S.~E.}\ \bibnamefont {Economou}}, \bibinfo
  {author} {\bibfnamefont {E.}~\bibnamefont {Barnes}}, \ and\ \bibinfo {author}
  {\bibfnamefont {N.~J.}\ \bibnamefont {Mayhall}},\ }\bibfield  {title}
  {\enquote {\bibinfo {title} {Is the trotterized uccsd ansatz chemically
  well-defined?}}\ }\href@noop {} {\bibfield  {journal} {\bibinfo  {journal}
  {J. Chem. Theory Comput.}\ }\textbf {\bibinfo {volume} {16}},\ \bibinfo
  {pages} {1--6} (\bibinfo {year} {2019}{\natexlab{b}})}\BibitemShut {NoStop}%
\bibitem [{\citenamefont {Sokolov}\ \emph {et~al.}(2020)\citenamefont
  {Sokolov}, \citenamefont {Barkoutsos}, \citenamefont {Ollitrault},
  \citenamefont {Greenberg}, \citenamefont {Rice}, \citenamefont {Pistoia},\
  and\ \citenamefont {Tavernelli}}]{sokolov2020qUCCSD}%
  \BibitemOpen
  \bibfield  {author} {\bibinfo {author} {\bibfnamefont {I.~O.}\ \bibnamefont
  {Sokolov}}, \bibinfo {author} {\bibfnamefont {P.~K.}\ \bibnamefont
  {Barkoutsos}}, \bibinfo {author} {\bibfnamefont {P.~J.}\ \bibnamefont
  {Ollitrault}}, \bibinfo {author} {\bibfnamefont {D.}~\bibnamefont
  {Greenberg}}, \bibinfo {author} {\bibfnamefont {J.}~\bibnamefont {Rice}},
  \bibinfo {author} {\bibfnamefont {M.}~\bibnamefont {Pistoia}}, \ and\
  \bibinfo {author} {\bibfnamefont {I.}~\bibnamefont {Tavernelli}},\ }\bibfield
   {title} {\enquote {\bibinfo {title} {Quantum orbital-optimized unitary
  coupled cluster methods in the strongly correlated regime: Can quantum
  algorithms outperform their classical equivalents?}}\ }\href {\doibase
  10.1063/1.5141835} {\bibfield  {journal} {\bibinfo  {journal} {J. Chem.
  Phys.}\ }\textbf {\bibinfo {volume} {152}},\ \bibinfo {pages} {124107}
  (\bibinfo {year} {2020})},\ \Eprint {http://arxiv.org/abs/1911.10864}
  {1911.10864} \BibitemShut {NoStop}%
\bibitem [{\citenamefont {Kim}\ \emph {et~al.}(2021)\citenamefont {Kim},
  \citenamefont {Wood}, \citenamefont {Yoder}, \citenamefont {Merkel},
  \citenamefont {Gambetta}, \citenamefont {Temme},\ and\ \citenamefont
  {Kandala}}]{youngseok2022zne}%
  \BibitemOpen
  \bibfield  {author} {\bibinfo {author} {\bibfnamefont {Y.}~\bibnamefont
  {Kim}}, \bibinfo {author} {\bibfnamefont {C.~J.}\ \bibnamefont {Wood}},
  \bibinfo {author} {\bibfnamefont {T.~J.}\ \bibnamefont {Yoder}}, \bibinfo
  {author} {\bibfnamefont {S.~T.}\ \bibnamefont {Merkel}}, \bibinfo {author}
  {\bibfnamefont {J.~M.}\ \bibnamefont {Gambetta}}, \bibinfo {author}
  {\bibfnamefont {K.}~\bibnamefont {Temme}}, \ and\ \bibinfo {author}
  {\bibfnamefont {A.}~\bibnamefont {Kandala}},\ }\href {\doibase
  10.48550/ARXIV.2108.09197} {\enquote {\bibinfo {title} {Scalable error
  mitigation for noisy quantum circuits produces competitive expectation
  values},}\ } (\bibinfo {year} {2021})\BibitemShut {NoStop}%
\bibitem [{\citenamefont {LaRose}\ \emph {et~al.}(2022)\citenamefont {LaRose},
  \citenamefont {Mari}, \citenamefont {Kaiser}, \citenamefont {Karalekas},
  \citenamefont {Alves}, \citenamefont {Czarnik}, \citenamefont {El~Mandouh},
  \citenamefont {Gordon}, \citenamefont {Hindy}, \citenamefont {Robertson},
  \citenamefont {Thakre}, \citenamefont {Wahl}, \citenamefont {Samuel},
  \citenamefont {Mistri}, \citenamefont {Tremblay}, \citenamefont {Gardner},
  \citenamefont {Stemen}, \citenamefont {Shammah},\ and\ \citenamefont
  {Zeng}}]{LaRose2022mitiqsoftware}%
  \BibitemOpen
  \bibfield  {author} {\bibinfo {author} {\bibfnamefont {R.}~\bibnamefont
  {LaRose}}, \bibinfo {author} {\bibfnamefont {A.}~\bibnamefont {Mari}},
  \bibinfo {author} {\bibfnamefont {S.}~\bibnamefont {Kaiser}}, \bibinfo
  {author} {\bibfnamefont {P.~J.}\ \bibnamefont {Karalekas}}, \bibinfo {author}
  {\bibfnamefont {A.~A.}\ \bibnamefont {Alves}}, \bibinfo {author}
  {\bibfnamefont {P.}~\bibnamefont {Czarnik}}, \bibinfo {author} {\bibfnamefont
  {M.}~\bibnamefont {El~Mandouh}}, \bibinfo {author} {\bibfnamefont {M.~H.}\
  \bibnamefont {Gordon}}, \bibinfo {author} {\bibfnamefont {Y.}~\bibnamefont
  {Hindy}}, \bibinfo {author} {\bibfnamefont {A.}~\bibnamefont {Robertson}},
  \bibinfo {author} {\bibfnamefont {P.}~\bibnamefont {Thakre}}, \bibinfo
  {author} {\bibfnamefont {M.}~\bibnamefont {Wahl}}, \bibinfo {author}
  {\bibfnamefont {D.}~\bibnamefont {Samuel}}, \bibinfo {author} {\bibfnamefont
  {R.}~\bibnamefont {Mistri}}, \bibinfo {author} {\bibfnamefont
  {M.}~\bibnamefont {Tremblay}}, \bibinfo {author} {\bibfnamefont
  {N.}~\bibnamefont {Gardner}}, \bibinfo {author} {\bibfnamefont {N.~T.}\
  \bibnamefont {Stemen}}, \bibinfo {author} {\bibfnamefont {N.}~\bibnamefont
  {Shammah}}, \ and\ \bibinfo {author} {\bibfnamefont {W.~J.}\ \bibnamefont
  {Zeng}},\ }\bibfield  {title} {\enquote {\bibinfo {title} {Mitiq: {A}
  software package for error mitigation on noisy quantum computers},}\ }\href
  {\doibase 10.22331/q-2022-08-11-774} {\bibfield  {journal} {\bibinfo
  {journal} {{Quantum}}\ }\textbf {\bibinfo {volume} {6}},\ \bibinfo {pages}
  {774} (\bibinfo {year} {2022})}\BibitemShut {NoStop}%
\end{thebibliography}

\end{document}